\documentclass[twocolumn]{aastex631}

\usepackage{mathrsfs}
\usepackage{ulem}
\usepackage{float}
\usepackage{amsmath}
\usepackage{rotating}
\usepackage{longtable}
\usepackage[T1]{fontenc}
\usepackage{soul}
\usepackage{float} 
\usepackage[title]{appendix} % Optional: For more control over the appendix

\shorttitle{Four planets around Barnard's Star}
\shortauthors{Basant et al.}
\begin{document}

\title{Four sub-Earth planets orbiting Barnard's Star from MAROON-X and ESPRESSO}

\correspondingauthor{Ritvik Basant}
\email{rbasant@uchicago.edu}

% rbasant@uchicago.edu (Lead author)
\author[0000-0003-4508-2436]{Ritvik Basant}
\affiliation{Department of Astronomy \& Astrophysics, University of Chicago, Chicago, IL 60637, USA}

% rluque@uchicago.edu (MAROON-X)
\author[0000-0002-4671-2957]{Rafael Luque}
\affiliation{Department of Astronomy \& Astrophysics, University of Chicago, Chicago, IL 60637, USA}
\affiliation{NHFP Sagan Fellow}

% jacobbean@uchicago.edu  (Advisor)
\author[0000-0003-4733-6532]{Jacob L.\ Bean}
\affiliation{Department of Astronomy \& Astrophysics, University of Chicago, Chicago, IL 60637, USA}

% seifahrt@uchicago.edu (MAROON-X instrument)
\author[0000-0003-4526-3747]{Andreas Seifahrt}
\affiliation{Gemini Observatory/NSF NOIRLab, 670 N. A'ohoku Place, Hilo, HI 96720, USA}

% MAROON-X)
\author[0000-0003-2404-2427]{Madison Brady}
\affiliation{Department of Astronomy \& Astrophysics, University of Chicago, Chicago, IL 60637, USA}

% MAROON-X)
\author[0000-0002-3852-3590]{Lily L.\ Zhao}
\affiliation{Department of Astronomy \& Astrophysics, University of Chicago, Chicago, IL 60637, USA}
\affiliation{NHFP Sagan Fellow}

% MAROON-X)
\author[0009-0003-1142-292X]{Nina Brown}
\affiliation{Department of Astronomy \& Astrophysics, University of Chicago, Chicago, IL 60637, USA}

% MAROON-X)
\author[0009-0005-1486-8374]{Tanya Das}
\affiliation{Department of Astronomy \& Astrophysics, University of Chicago, Chicago, IL 60637, USA}

\author[0000-0002-4410-4712]{Julian St{\"u}rmer}
\affiliation{Landessternwarte, Zentrum f{\"u}r Astronomie der Universität Heidelberg, K{\"o}nigstuhl 12, D-69117 Heidelberg, Germany}

\author[0000-0003-0534-6388]{David Kasper}
\affiliation{Department of Astronomy \& Astrophysics, University of Chicago, Chicago, IL 60637, USA}

\author[0009-0002-1141-5853]{Rohan Gupta}
\affiliation{Department of Astronomy \& Astrophysics, University of Chicago, Chicago, IL 60637, USA}

\author[0000-0001-7409-5688]{Guðmundur Stefánsson} 
\affil{Anton Pannekoek Institute for Astronomy, University of Amsterdam, Science Park 904, 1098 XH Amsterdam, The Netherlands} 

\begin{abstract}
Barnard's Star is an old, single M dwarf star that comprises the second-closest extrasolar system. It has a long history of claimed planet detections from both radial velocities and astrometry. However, none of these claimed detections have so far withstood further scrutiny. Continuing this story, extreme precision radial velocity (EPRV) measurements from the ESPRESSO instrument have recently been used to identify four new sub-Earth-mass planet candidates around Barnard's Star. We present here 112 radial velocities of Barnard's Star from the MAROON-X instrument that were obtained independently to search for planets around this compelling object. The data have a typical precision of 30\,cm\,s$^{-1}$ and are contemporaneous with the published ESPRESSO measurements (2021 -- 2023). The MAROON-X data on their own confirm planet b ($P$\,=\,3.154\,d) and planet candidates c and d ($P$\,=\,4.124\,d and 2.340\,d, respectively). Furthermore, adding the MAROON-X data to the ESPRESSO data strengthens the evidence for planet candidate e ($P$\,=\,6.739\,d), thus leading to its confirmation. The signals from all four planets are $<$50\,cm\,s$^{-1}$, the minimum masses of the planets range from 0.19 to 0.34\,$M_{\oplus}$, and the system is among the most compact known among late M dwarfs hosting low-mass planets. The current data rule out planets with masses $>0.57\,M_{\oplus}$ (with a $99\%$ detection probability) in Barnard Star's habitable zone ($P$\,=\,10 -- 42\,d).
\end{abstract}

%%Continued observations of Barnard's Star and other nearby stars with EPRV instruments like MAROON-X and ESPRESSO hold significant promise for additional exoplanet discoveries.

%\keywords{Exoplanets (498), Exoplanet systems (484), Radial velocity (1332), M dwarf stars (982)}


\section{Introduction} \label{sec:intro}
At a distance of only 1.83\,pc \citep{gaia21}, Barnard's Star is a touchstone for studies of nearby stars. It is a single, mid-M dwarf star that is notable for having the largest proper motion of any star in the sky. Barnard's Star is thought to be $\sim$10\,Gyr old due to its slow rotation and low levels of activity (see discussions in \citealt{gauza15} and \citealt{france20}). An up-to-date summary of Barnard Star's properties can be found in Table 1 of \citet{gonzalez24}. Note that another common name for Barnard's Star is Gl\,699 \citep{gliese57}.

Recently, radial velocity observations using the EPRV spectrograph ESPRESSO on the VLT \citep{pepe21} have revealed four very low-mass planet candidates in close-in orbits ($P$ = 2 -- 7\,d) around Barnard's Star \citep{gonzalez24}. The detected signals have semi-amplitudes that range from 47 down to 20\,cm\,s$^{-1}$, and they represent some of the smallest radial velocity signals claimed to date. \citet{gonzalez24} were confident enough to declare the signal at $P$\,=\,3.154\,d as due to a bona-fide planet (``Barnard b''), but they left the other signals as planet candidates.

Due to its nearness and low mass, Barnard's Star has been the subject of many searches for planets. Although there have been claims of planet detections using both astrometry and radial velocities dating back decades, none of the previous claimed detections survived additional observations and analyses \citep{gatewood95,benedict99,lubin21,artigau22}. In an attempt to confirm and measure the radius of Barnard b, \citet{stefanov24} looked for transits in TESS data, but they did not detect any transit signals.

In this paper, we present an extensive set of radial velocity measurements of Barnard's Star from the MAROON-X instrument on Gemini-N. MAROON-X is an EPRV spectrograph with a wavelength coverage (500 -- 920\,nm) optimized for M dwarfs like Barnard's Star \citep{2016SPIE.9908E..18S, 2018SPIE10702E..6DS, 2020SPIE11447E..1FS, 2022SPIE12184E..1GS}. We describe our MAROON-X observations and data reduction in \S\ref{sec:observations_data_reduction}. We present analyses of the MAROON-X and ESPRESSO data that confirm and extend the results of \citet{gonzalez24} in \S\ref{sec:analysis}. We conclude with a discussion in \S\ref{sec:discussion}.

\section{Observations and Data Reduction} \label{sec:observations_data_reduction}
We have been observing Barnard's Star with MAROON-X regularly since 2021 because it is a key target of our ongoing search for planets around stars within 4\,pc. We have previously published results from our 4\,pc planet search for Wolf 359 and Teegarden's Star \citep{bowens-rubin23, dreizler24}. We observed Barnard's Star as a regular science target and we also observed it as a calibration target to validate the performance of MAROON-X for other programs. The data presented in this paper cover three seasons from 2021 to 2023. MAROON-X was used in campaign mode during this period, with discrete observing runs ranging from one to five weeks in duration. The Barnard's Star observations we analyze here were obtained during nine distinct observing runs.

All the MAROON-X observations of Barnard's Star were five minute exposures, and most visits were comprised of just a single exposure. Barnard's Star is so bright for Gemini+MAROON-X that we mostly observed it in mid- to below-average conditions (i.e., 85th percentile image quality and 70th percentile cloud cover were typical constraints). We even occasionally observed it in very poor conditions as part of Gemini's ``Band 4'' time (i.e., heavy cloud cover and/or bad seeing). These poor weather observations typically had multiple successive exposures because otherwise, the telescope would have been idle. All the individual exposures were reduced separately. We discarded seven exposures that were taken in particularly bad observing conditions and thus have anomalously low signal-to-noise. We then binned the radial velocity values for observations taken within 60 minutes. The total data set used here is based on 151 individual exposures obtained over 112 unique epochs. 

We reduced and extracted wavelength-calibrated 1D spectra from Barnard's Star data using our standard pipeline, which was first used for radial velocity measurements and described by \citet{trifonov21} and \citet{winters22}. We measured radial velocities from the reduced spectra using a specific version of the \texttt{serval} package \citep{2018A&A...609A..12Z} that was modified to work with MAROON-X data. The \texttt{serval} code uses a template-matching approach where the observed spectrum with the highest signal-to-noise ratio (SNR) is chosen as a reference. The remaining spectra are shifted to this reference frame and then \texttt{serval} co-adds them to produce a high-SNR template. The template-matching approach for computing radial velocities is more suitable for the rich spectra of M dwarfs than cross-correlation with binary templates. The telluric mask provided within \texttt{serval} was manually checked against a spectrum of an extremely fast rotating ($v\sin i \sim 100$\,km\,s$^{-1}$) A0-type star obtained via MAROON-X in fairly wet conditions for Maunakea (precipitable water vapor $\sim 5$\,mm) at 1.5 airmass. This custom mask contains all lines deeper than $1\%$ and is not dependent on airmass or precipitable water vapor.

In addition to radial velocities, \texttt{serval} measures various activity indicators, including the chromatic, differential line width, H$\alpha$, Ca II triplet, and Na I doublet indices. MAROON-X has two arms, ``Blue'' ($\lambda$\,=\,500 -- 670\,nm) and ``Red'' ($\lambda$\,=\,650 -- 920\,nm), which are reduced and analyzed separately. We used the \texttt{barycorrpy} code \citep{kanodia18} to calculate the barycentric correction, including the effect of secular acceleration. The barycentric Earth radial velocities for the MAROON-X observations range from $-24.825$ to $+26.526$\,km\,s$^{-1}$. The radial velocities are given in Table~\ref{table:dataset}.

MAROON-X has been using a passively-stabilized Fabry-P\'erot etalon \citep{2017JATIS...3b5003S} as the primary calibration source since its first light. Simultaneous etalon spectra are obtained using a dedicated fiber during each science exposure. The simultaneous etalon spectra are used to measure the drift of the instrument from reference etalon spectra taken through the science fibers during daytime calibrations. The etalon parameters were determined once in 2020 by comparison to a ThAr reference spectrum. The etalon is observed to drift at the low level of 2\,cm\,s$^{-1}$ per day based on comparison with ThAr spectra taken between August 2021 and October 2023 \citep{2025arXiv250215074B}. As described in the next section, we correct this drift using free offsets between observing runs rather than applying a correction from the ThAr data or from the ensemble approach of \cite{2025arXiv250215074B}.

The reduced MAROON-X spectra for Barnard's Star have a peak SNR of 176 and 460 per pixel in the wavelength ranges of 633 -- 643 and 800 -- 812\,nm for the Blue and Red arms, respectively. The median radial velocity errors from \texttt{serval} are 37\,cm\,s$^{-1}$ for Blue arm data, and 29\,cm\,s$^{-1}$ for Red arm data. For comparison, the ESPRESSO data from \citet{gonzalez24} have median errors of 10\,cm\,s$^{-1}$ based on spectra taken with three times longer exposures (15\,minutes vs.\ our 5\,minutes) and utilizing the full wavelength range of the data. Taking the weighted mean of our Blue and Red arm data and dividing the error by $\sqrt{3}$ gives 13\,cm\,s$^{-1}$, which provides a comparison to the ESPRESSO precision. MAROON-X's throughput is lower than ESPRESSO's, partially due to a smaller fiber entrance on the sky (0.77\arcsec\ vs.\ 1.0\arcsec), and we rarely observed Barnard's Star in good conditions. ESPRESSO also has a higher spectral resolution ($R \sim$ 138,000 vs.\ 85,000), which is an advantage for radial velocity measurements. However, MAROON-X has a redder wavelength coverage (red cutoff of 920 vs.\ 788\,nm) that is better for M dwarfs \citep{Reiners_2020}.

\section{Methods and Results}\label{sec:analysis}

\subsection{Modeling Stellar Activity}\label{subsec:stellar-activity}
Besides a long-term magnetic activity cycle with $P\sim3800$\,d \citep[][]{2019MNRAS.488.5145T}, radial velocities of Barnard's Star exhibit activity-induced variation at both the rotational period ($\sim142$\,d) and its second harmonic \citep[$\sim71$\,d,][]{gonzalez24}. Such stellar activity-induced radial velocity variations not only produce statistically significant signals in the periodograms \citep{2001A&A...379..279Q}, but also pose a significant challenge in characterizing sub-\,m\,s$^{-1}$ signals \citep[][]{2016PASP..128f6001F}. 

Gaussian Processes (GP) have been proven to be an efficient way to mitigate the effects of stellar activity in the radial velocity data \citep[see e.g.,][]{2015MNRAS.452.2269R}. \citet{gonzalez24} performed a detailed analysis of the activity-induced variations in Barnard's Star radial velocities by modeling them simultaneously along with the Full-Width at Half Maximum (FWHM) of the CCF. The authors concluded that a Double Simple Harmonic Oscillator (DSHO) kernel, centered on the rotational period and its second harmonic, can effectively model the activity signals in data for Barnard's Star. 

For an initial test, we compute the generalized Lomb-Scargle (GLS) periodogram \citep{2009A&A...496..577Z} of the raw MAROON-X radial velocities (shown in the top panel of Figure~\ref{fig:activity} in the Appendix), provided via PyAstronomy \citep{2019ascl.soft06010C}. The raw radial velocities exhibit periodic signals associated with the stellar rotation period ($\sim 142$ days; \citet{gonzalez24}). Moreover, a prominent signal near the edge of the frequency grid ($\sim 600 d$) is observed, which we associate with the MAROON-X instrumental drift.

As an additional diagnostic, we computed periodograms for the spectroscopic activity indicators, including the chromatic index, differential line width, and $H-\alpha$ index (see the bottom panels of Figure~\ref{fig:activity} in the Appendix). The chromatic index measures the correlation between the radial velocities and wavelength, while the differential line width characterizes width changes in individual line profiles, serving as an analog to the FWHM in the cross-correlation technique. The periodogram of the chromatic index from the Red channel data of MAROON-X exhibits significant signals around both the stellar rotation period and its second harmonic. Similarly, the periodogram of the differential line width from the Red channel data of MAROON-X shows significant signals around the second harmonic of the rotation period, within the error bars. Although the differential line width of the Blue channel in the MAROON-X data exhibits similar trends, the chromatic index does not. We attribute this difference to the variation in signal-to-noise between the two arms; the Red arm data are twice as precise as the Blue arm data due to differing amounts of radial velocity information content in the respective spectra for this red star.

Consequently, as our analysis of the stellar activity is consistent with the results presented in \cite{gonzalez24}, we decided to adopt a DSHO GP kernel. We use \texttt{juliet} \citep{2019MNRAS.490.2262E}, a versatile modeling tool that builds upon \texttt{RadVel} \citep{2018PASP..130d4504F}, to model the radial velocity data. The \texttt{juliet} code employs \texttt{dynesty} \citep{2020MNRAS.493.3132S} to perform nested sampling and estimates the (log-)marginal likelihood ($\ln Z$), a useful statistical measure to compare models. 

The DSHO GP kernel, available within \texttt{juliet} via \texttt{celerite2} \citep{2017AJ....154..220F, 2018RNAAS...2...31F}, is parameterized by $\sigma_{\rm GP-INST}$ (standard deviation of the data), $\rm P_{rot}$ (rotational period of the primary oscillator), $Q_{0}$ (quality factor for the second oscillator), $f$ (ratio of amplitudes of the secondary and the primary mode), and lastly, $dQ$ (the difference between the quality factors of both oscillators). We note that we do not use a 2-dimensional GP model as was done by \cite{gonzalez24}, but instead we only model the radial velocities. Although much simpler, it can still accurately recover the planet parameters (well within $1\sigma$) derived by \cite{gonzalez24} for both the 1-planet and 4-planet models (see Figure~\ref{fig:phase-folded-esp} in the Appendix).

\begin{figure*}[ht!]
    \centering
    \includegraphics[width=1.5\columnwidth,keepaspectratio]{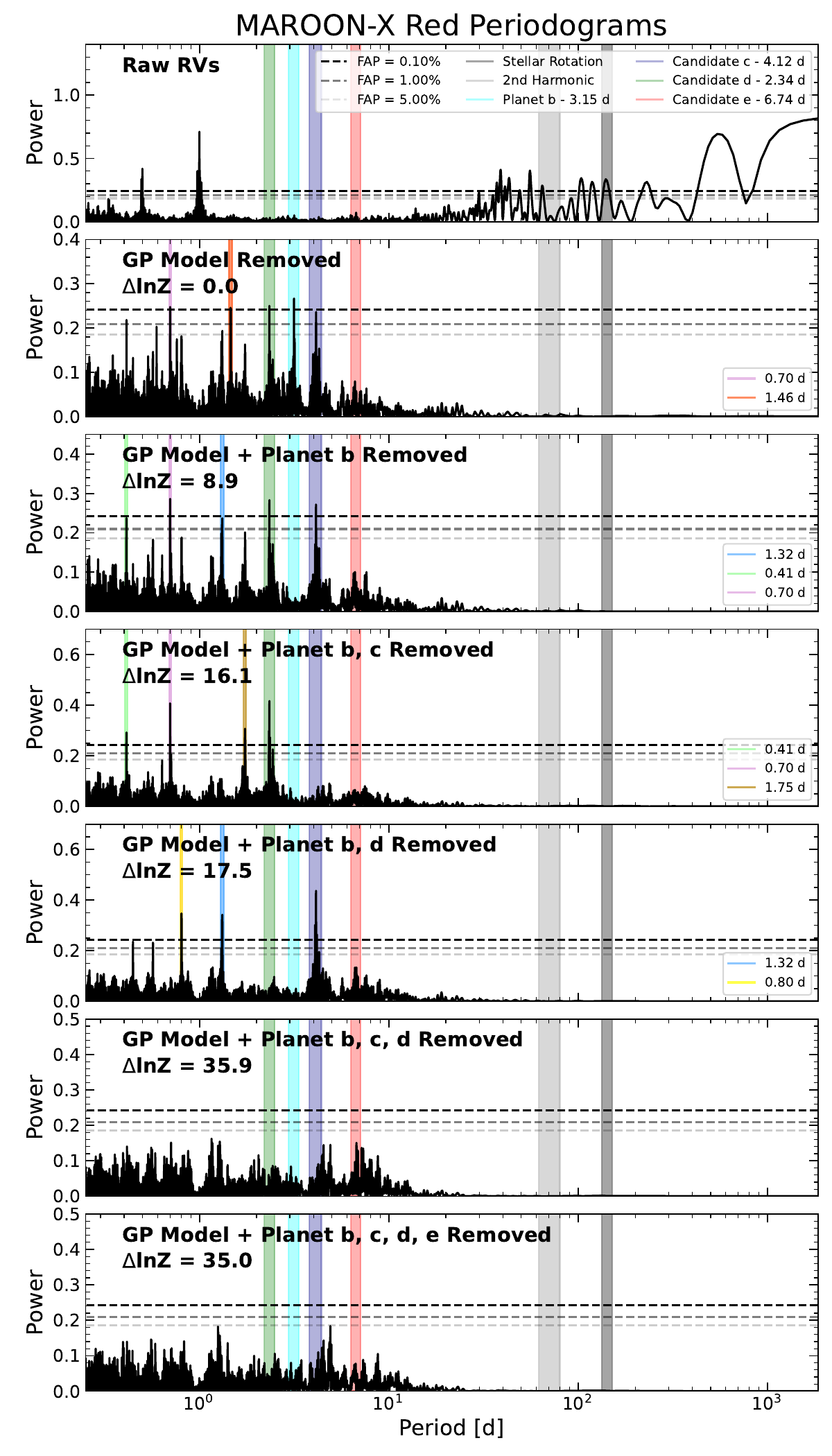} % Adjust the size to fit within the page
    \caption{This plot shows the computed GLS periodograms for the residuals of MAROON-X Red channel radial velocities. A beta distribution for the planetary eccentricities has been used for all the planets. }
    \label{fig:periodogram-mxred}
\end{figure*}

In addition to the GP model, we also model an additional white noise (jitter) term ($\sigma$) in the radial velocity data. Lastly, we elected to use free offsets ($\gamma$) between the MAROON-X observing runs because this approach has proven highly effective for probing short-period ($P \lesssim$\,15\,d) planets like the Barnard's Star candidates. For example, MAROON-X data with free run offsets have been used to measure the velocity semi-amplitudes of close-in TESS planets to 10\,cm\,s$^{-1}$ or better \citep[e.g.,][]{caballero22,brady24}. Repeating the analysis with run offsets constrained by an ensemble analysis of many MAROON-X targets \citep{2025arXiv250215074B} gives similar results.

GP models are very flexible, which could lead to data overfitting. To prevent this, we use a normal prior on the rotation period hyperparameter ($\rm P_{rot}$) of the GP with the mean 142 d and standard deviation 9 d as derived in \cite{gonzalez24}. We use a uniform prior ($\mathcal{U}$ [0, 1]) on hyperparameter $f$ and a log-uniform prior $\mathcal{LU}$ [0.1, 100]) on both $Q_{0}$ and $dQ$. We share the GP hyperparameters $f$, $dQ$, $Q_{0}$, and $\rm P_{rot}$ across different runs because of two reasons: (1) these hyperparameters are related to the time-dependent properties of the oscillator and are less likely to be wavelength-dependent for radial velocity signals caused by stellar surface inhomogeneities, and (2) sharing these hyperparameters between runs helps in more efficient mitigation of the stellar activity thanks to the longer baseline to constrain these characteristic parameters. The GP hyperparameter $\sigma_{\rm GP-INST}$ is directly related to the amplitude of the radial velocity signal imprinted by the stellar activity and is, therefore, wavelength-dependent \cite[][]{huelmo+08, reiners10, crockett+10}. Therefore, we chose not to share $\sigma_{\rm GP-INST}$ between different instruments, but only between multiple datasets from the same instrument. It is assigned a uniform prior $\mathcal{U}$ [0, 10].

\subsection{Independent detection of planets b, c, and d using MAROON-X}\label{subsec:mx-independent}

%\begin{figure}[]
%    \centering
%    \includegraphics[width=1\columnwidth,keepaspectratio]{MX-GP-residuals.pdf} % Adjust the size to fit within the page
%    \caption{The top panel in this plot shows the residuals of MAROON-X Red channel radial velocities after removing the GP model. The bottom panel shows the GLS periodogram of the residuals shown in the top panel.}
%    \label{fig:mxred-residuals-0p}
%\end{figure}

\begin{table}[]
    \centering
    \begin{tabular}{ccc}
    \hline
    \hline
    Model & $\ln Z$ & $\Delta\ln Z$ \\    
    \hline    
    GP-only Model & -183.5 & 0\\
    \hline
    \multicolumn{3}{c}{Circular Orbit}\\
    \hline
    1-Planet (b) & -174.1 & 9.4 \\
    2-Planet (b, c) & -169.6 & 14.0 \\
    2-Planet (b, d) & -164.3 & 19.2 \\
    3-Planet (b, c, d) & -147.9 & 35.6 \\
    \hline    
    \multicolumn{3}{c}{Beta Prior on Eccentricity}\\
    \hline
    1-Planet (b) & -174.6 & 8.9 \\
    2-Planet (b, c) & -167.5 & 16.1 \\
    2-Planet (b, d) & -166.1 & 17.5 \\
    \textbf{3-Planet (b, c, d)} & \textbf{-147.6} & \textbf{35.9} \\
    4-Planet (b, c, d, e) & -148.5 & 35.0\\
    \hline    
    \multicolumn{3}{c}{Uninformative Prior on Eccentricity}\\
    \hline
    1-Planet (b) & -176.7 & 6.9 \\
    2-Planet (b, c) & -169.5 & 14 \\
    2-Planet (b, d) & -166.7 & 16.9 \\
    3-Planet (b, c, d) & -144.6 & 38.9 \\
    \hline
    \end{tabular}
    \caption{This table summarizes the marginal evidence for different models fitted to the MAROON-X Red channel radial velocities. The 3-Planet model assuming the $\beta$ prior on the eccentricities (bold) is the preferred fit to these data.}
    \label{tab:lnz-mx}
\end{table}

\begin{table}[]
    \centering
    \begin{tabular}{ccc}
    \hline
    \hline
    Model & $\ln Z$ &  $\Delta\ln Z$ \\    
    \hline
    \multicolumn{3}{c}{Global GP}\\
    \hline
    GP-only Model & -412.3 & 0\\
    1-Planet (b) & -386.6 & 25.7\\
    2-Planet (b, d) & -365.2 & 47.1 \\
    3-Planet (b, c, d) & -333.7 & 78.6 \\
    {4-Planet (b, c, d, e)} & {-326.4} & {85.9} \\
    \hline    
    \multicolumn{3}{c}{Instrument-by-Instrument GP}\\
    \hline
    GP-only Model & -419.8 & 0\\
    1-Planet (b) & -393.0 & 26.8\\
    2-Planet (b, d) & -370.3 & 49.5 \\
    3-Planet (b, c, d) & -332.8 & 87.0 \\
    \textbf{4-Planet (b, c, d, e)} & \textbf{-325.6} & \textbf{94.3} \\
    \hline    
    \end{tabular}
    \caption{This table summarizes the change in marginal evidence between different models for the MAROON-X Red channel + ESPRESSO data assuming the $\beta$ prior on the eccentricities.}
    \label{tab:lnz-combined}
\end{table}

We first analyzed the MAROON-X data on their own to provide an independent assessment of the planet candidates proposed by \citet{gonzalez24}. We focused our primary analyses on the Red arm data from MAROON-X because they are potentially less influenced by stellar activity. The results are all consistent or even better (e.g., higher marginal evidence and smaller parameter uncertainties) when including the Blue arm data. Figures showing the results of analyses of the Blue arm data (Figures~\ref{fig:phase-folded-rb}, \ref{fig:periodogram-rb}, \ref{fig:periodogram-mxrb-esp}, and \ref{fig:phase-folded-mxrb-esp}) can be found in the Appendix. Ultimately, we find that the root mean square (RMS) of the residuals is about twice as large in the Blue arm data compared to the Red arm data despite the photon-limited errors only being 27\% larger.

The second panel of Figure~\ref{fig:periodogram-mxred} shows the residuals after removing the GP model from the MAROON-X Red channel radial velocities. The periodogram shows 2 significant signals (false alarm probability FAP $> 0.1 \%$) at the locations of planet b and candidate d, and a third signal (FAP $>$ $1\%$) at the location of planet candidate c. 

To characterize these signals, we follow an iterative strategy. Once we identify a significant signal having FAP $> 0.1 \%$, we model it in the radial velocity data using a Keplerian orbit. We then compute the residuals and recompute the periodogram. In the case of multiple significant signals, we model only one of them and check if the other signals still appear in the periodogram and then repeat this process for all other signals. We keep adding additional Keplerian signals until the periodogram of the residuals does not show any significant signals. We ran similar models multiple times to analyze possible variances in the marginal evidence. Finally, following \cite{2008ConPh..49...71T}, we confirm a signal originating from an orbiting planet if the improvement in the marginal evidence is $> 5.0$. 

The periodograms with successive planets removed are shown in Figure~\ref{fig:periodogram-mxred}. The periodograms show multiple high-frequency signals with FAP $\ga$ 1\% in addition to those we identify as due to planets. We associate them with the following aliases of the planetary signals: $0.70$ d (1-d alias of $2.34$ d signal), $0.41$ d (1-d alias of 0.70 d signal), $1.75$ d (1-d alias of $2.34$ d signal), $1.46$ d (1-d alias of $3.15$ d signal), $1.32$ d (1-d alias of $4.12$ d signal), and $0.80$ d (1-d alias of $4.12$ d signal).

The Keplerian orbits in \texttt{juliet} were parameterized using orbital period ($P$), velocity semi-amplitude ($K$), and time of periastron passage ($t_{0}$). Given the high computational cost associated with modeling a large number of parameters, we decided to use a uniform prior on the planetary period with a width of $0.1$ d centered around the values estimated via periodogram analysis. Furthermore, as the semi-amplitudes for all four planet candidates around Barnard's Star are less $\sim 0.5$\,m\,s$^{-1}$, we used a uniform prior on semi-amplitude $\mathcal{U}$ $[0, 2]$ \,m\,s$^{-1}$.

To select the best parameterization for the eccentricities, we tested three different scenarios: (1) Parametrizing the eccentricities as $h = \sqrt{e}\sin\omega$ and $k = \sqrt{e}\cos\omega$ and using an uninformative uniform prior $\mathcal{U}$ [-1, 1] on both these parameters; (2) Drawing eccentricities directly from a beta distribution $\mathcal{\beta}$ [1.52, 29] and drawing the argument of periastron ($\omega$) from a uniform distribution $\mathcal{U}$ [-180, 180]. This eccentricity distribution is well-suited for systems with multiple transiting planets \citet{2019AJ....157...61V}; and lastly (3) Assuming circular orbits.

\begin{figure*}[]
    \centering
    \includegraphics[width=0.8\textwidth, keepaspectratio]{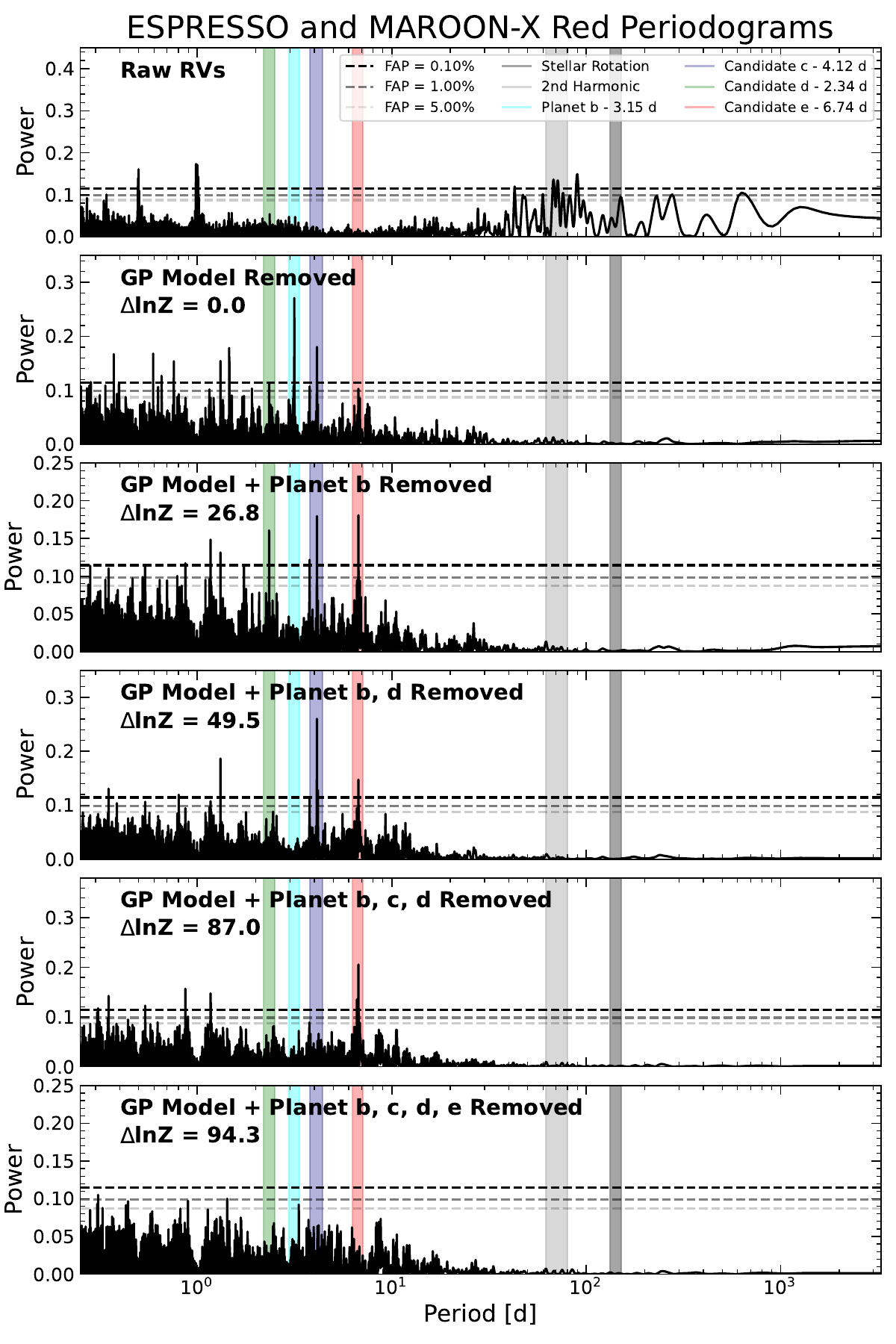} % Adjust the size to fit within the page
    \caption{This plot shows the computed GLS periodograms for the combined residuals of MAROON-X Red channel and ESPRESSO radial velocities. A beta distribution for the planetary eccentricities has been used for all planets. }
    \label{fig:periodogram-combined}
\end{figure*}

The results of these tests are presented in Table~\ref{tab:lnz-mx}. In all cases, the marginal evidence improves by $> 5.0$ over the previous model. However, our results from using an uninformative prior on the eccentricity parameters $h$ and $k$ results in high eccentricity solutions ($0.10^{+0.12}_{-0.07}$ for planet b, $0.24^{+0.11}_{-0.11}$ for planet c, and $0.44^{+0.09}_{-0.10}$ for planet d), which likely arise due to the low signal-to-noise of the data. Such high eccentricity solutions yield marginal evidences that are comparable to the low eccentricity solutions (e.g., eccentricities drawn from a $\beta$-distribution). Additionally, such high eccentricities will likely make the system unstable (see \S~\ref{subsec:stability}). Drawing eccentricities directly from a $\mathcal{\beta}$-distribution forces the eccentricities to be low, which is typical for compact multi-planet systems \citep{2019AJ....157...61V}, while still exploring high eccentricity solutions. Moreover, unlike the case of circular orbits, this parameterization of eccentricity preserves some information about the possible eccentricities of these planets. Therefore, we decided to draw eccentricities from a $\mathcal{\beta}$ [1.52, 29] distribution and the argument of periastron from a uniform $\mathcal{U}$ [-180, 180] distribution for all models used in the remainder of this paper.

\begin{figure*}[]
    \centering
    \includegraphics[width=0.9\textwidth,height=1.0\textheight,keepaspectratio]{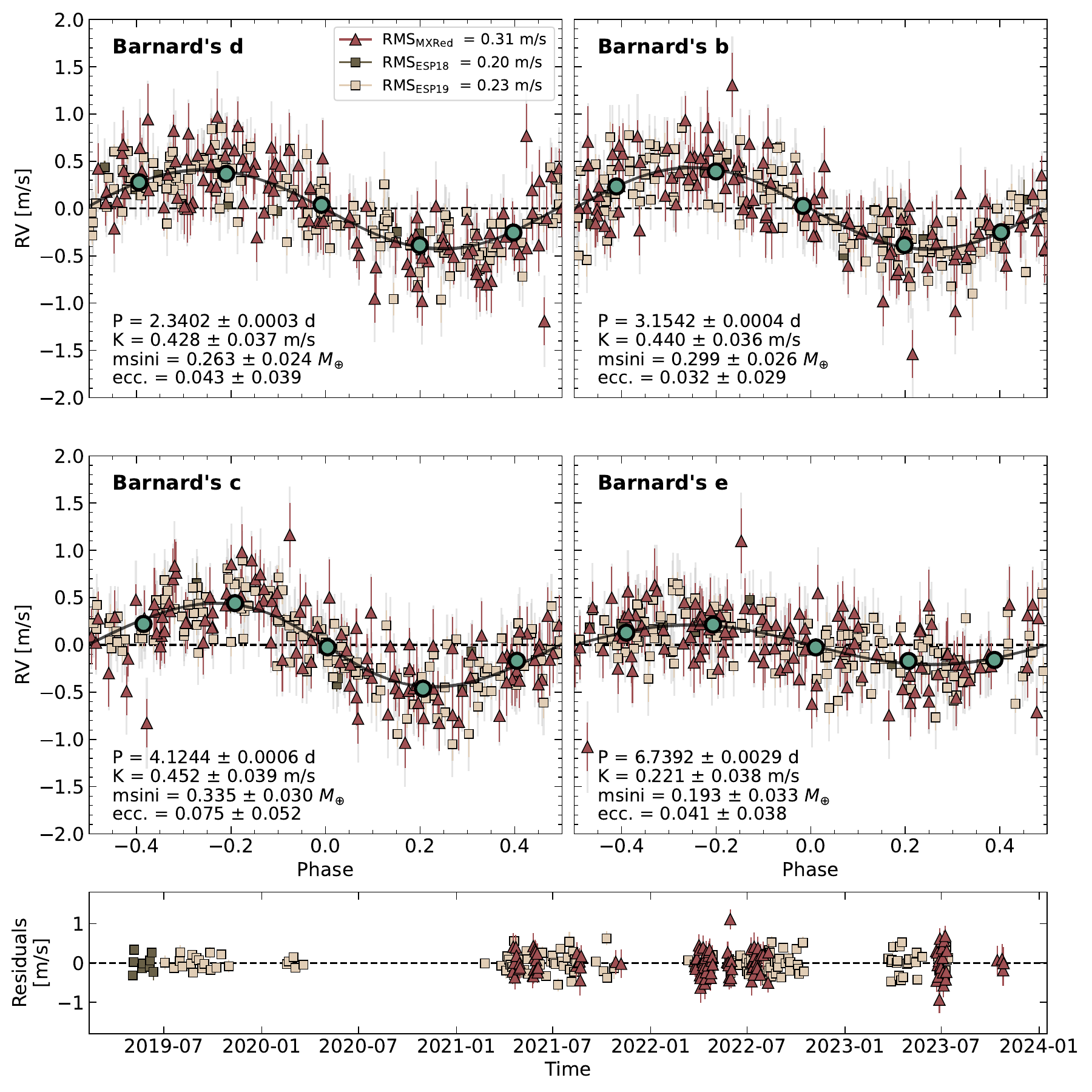} % Adjust the size to fit within the page
    \caption{Top panels: Phase-folded plots for planets Barnard b, c, d, and e based on joint fit between MAROON-X Red channel and ESPRESSO radial velocities. Bottom panel: Residuals as a function of time for the 4-Planet model.}
    \label{fig:phase-folded-combined}
\end{figure*}

We began by modeling the most significant signal at 3.154\,d, which corresponds to planet b. Once this signal is modeled and removed from the radial velocity data, the periodogram of the residuals shows two significant signals at the locations of planet candidates c ($P$\,=\,4.124\,d) and d ($P$\,=\,2.340\,d, see the third panel in Figure~\ref{fig:periodogram-mxred}). Additionally, for the 1-Planet model, the marginal evidence improves significantly over the GP-only model by 8.9. 

We then modeled both of the additional signals, one by one. The periodograms of the residuals of the 2-Planet models each show the third, unmodeled signal (shown in the fourth and fifth panels of Figure~\ref{fig:periodogram-mxred}). Irrespective of the order in which the second planet is added, the marginal evidence of the 2-Planet model improves significantly over the 1-Planet model (see Table~\ref{tab:lnz-mx}).

\begin{table*}[]
    \centering
    \begin{tabular}{ccc}
        \hline
        \hline
        \multicolumn{3}{c}{Planetary Parameters} \\
        \hline
        Parameter & Posterior & Prior Distribution \\
        \hline
        $\rm P_{d}$ [d] & $2.3402_{-0.0003}^{+0.0003}$ & $\mathcal{U}$ $[2.29, 2.39]$ \\ 
        $\rm t_{0, d}$ [BJD - 2450000] & $10243.70_{-0.07}^{+0.08}$ & $\mathcal{U}$ $[10242.54, 10244.89]$ \\ 
        $\rm K_{d}$ [m\,s$^{-1}$] & $0.428_{-0.036}^{+0.036}$ & $\mathcal{U}$ $[0, 2]$ \\ 
        $\rm e_{d}$  & $0.04_{-0.03}^{+0.05}$ & $\mathcal{\beta}$ $[1.52, 29.0]$ \\ 
        $\rm \omega_{d}$ [$^{\circ}$] & $-51.8_{-93.5}^{+190.8}$ & $\mathcal{U}$ $[-180, 180]$ \\ 
        $\rm m_{d}\sin i_{d}$ [$M_{\oplus}$] & $0.263 \pm 0.024$ & - \\
        $\rm a_{d}$ [au] & $0.0188 \pm 0.0003$ & - \\
        $\rm T_{\rm eq, d}$ [K]\tablenotemark{a} & $483$ & - \\
        \hline
        $\rm P_{b}$ [d] & $3.1542_{-0.0004}^{+0.0004}$ & $\mathcal{U}$ $[3.1, 3.2]$ \\ 
        $\rm t_{0, b}$ [BJD - 2450000] & $10243.38_{-0.09}^{+0.09}$ & $\mathcal{U}$ $[10242.12, 10245.32]$ \\ 
        $\rm K_{b}$ [m\,s$^{-1}$] & $0.440_{-0.036}^{+0.036}$ & $\mathcal{U}$ $[0, 2]$ \\ 
        $\rm e_{b}$  & $0.03_{-0.02}^{+0.03}$ & $\mathcal{\beta}$ $[1.52, 29.0]$ \\ 
        $\rm \omega_{b}$ [$^{\circ}$] & $+3.8_{-121.4}^{+117.7}$ & $\mathcal{U}$ $[-180, 180]$ \\ 
        $\rm m_{b}\sin i_{b}$ [$M_{\oplus}$] & $0.299 \pm 0.026$ & - \\
        $\rm a_{b}$ [au] & $0.0229 \pm 0.0003$ & - \\
        $\rm T_{\rm eq, b}$ [K]\tablenotemark{a} & $438$ & - \\
        \hline
        $\rm P_{c}$ [d] & $4.1244_{-0.0006}^{+0.0006}$ & $\mathcal{U}$ $[4.07, 4.17]$ \\ 
        $\rm t_{0, c}$ [BJD - 2450000] & $10242.92_{-0.10}^{+0.10}$ & $\mathcal{U}$ $[10241.62, 10245.82]$ \\ 
        $\rm K_{c}$ [m\,s$^{-1}$] & $0.452_{-0.038}^{+0.038}$ & $\mathcal{U}$ $[0, 2]$ \\ 
        $\rm e_{c}$  & $0.08_{-0.05}^{+0.06}$ & $\mathcal{\beta}$ $[1.52, 29.0]$ \\ 
        $\rm \omega_{c}$ [$^{\circ}$] & $90.8_{-48.1}^{+38.9}$ & $\mathcal{U}$ $[-180, 180]$ \\ 
        $\rm m_{c}\sin i_{c}$ [$M_{\oplus}$] & $0.335 \pm 0.030$ & - \\
        $\rm a_{c}$ [au] & $0.0274 \pm 0.0004$ & - \\
        $\rm T_{\rm eq, c}$ [K]\tablenotemark{a} & $400$ & - \\
        \hline
        $\rm P_{e}$ [d] & $6.7392_{-0.0028}^{+0.0028}$ & $\mathcal{U}$ $[6.69, 6.79]$ \\ 
        $\rm t_{0, e}$ [BJD - 2450000] & $10245.30_{-0.36}^{+0.37}$ & $\mathcal{U}$ $[10240.32, 10247.12]$ \\ 
        $\rm K_{e}$ [m\,s$^{-1}$] & $0.221_{-0.037}^{+0.037}$ & $\mathcal{U}$ $[0, 2]$ \\ 
        $\rm e_{e}$  & $0.04_{-0.03}^{+0.04}$ & $\mathcal{\beta}$ $[1.52, 29.0]$ \\ 
        $\rm \omega_{e}$ [$^{\circ}$] & $-27.5_{-96.1}^{+137.5}$ & $\mathcal{U}$ $[-180, 180]$ \\ 
        $\rm m_{e}\sin i_{e}$ [$M_{\oplus}$] & $0.193 \pm 0.033$ & - \\
        $\rm a_{e}$ [au] & $0.0381 \pm 0.0005$ & - \\
        $\rm T_{{\rm eq}, e}$ [K]\tablenotemark{a} & $340$ & - \\
        \hline
        \multicolumn{3}{c}{GP Hyperparameters}\\
        \hline
        $\sigma_{\rm GP-ESP}$ & $1.98_{-0.22}^{+0.28}$ & $\mathcal{U}$ $[0, 10]$\\
        $\sigma_{\rm GP-MXRed}$ & $1.78_{-0.33}^{+0.45}$ & $\mathcal{U}$ $[0, 10]$\\
        GP$\rm -P_{rot}$ & $140.99_{-8.41}^{+8.27}$ & $\mathcal{N}$ $[142, 9]$\\
        GP$\rm -Q0$ & $0.37_{-0.21}^{+0.39}$ & $\mathcal{LU}$ $[0.1, 6]$\\
        GP$\rm -f$ & $0.72_{-0.30}^{+0.20}$ & $\mathcal{U}$ $[0, 1]$\\
        GP$\rm -dQ$ & $0.25_{-0.12}^{+0.44}$ & $\mathcal{LU}$ $[0.1, 25]$\\
        \hline
        \end{tabular}
    \tablenotetext{a}{$T_{eq}$ is calculated assuming zero albedo and full heat redistribution}
    \caption{This table summarizes the posteriors and the priors for the four planets as derived from a joint fit between MAROON-X Red channel and ESPRESSO radial velocities.}
     \label{tab:planets+GP-combined}
\end{table*}

Finally, we added a third Keplerian signal in our model. We ran this model five times to analyze any variance in marginal evidence. The $\ln Z$ ranges from [$-149.7$, $-146.2$] with a median and standard deviation of $-147.6\pm1.2$. Although the posteriors for all runs are consistent, the runs with the lowest $\ln Z$ have a few samples drawn around the 1-year aliases of planets c and d. We identify this as the main source of variance in $\ln Z$. We report the median values of $\ln Z$ in Table~\ref{tab:lnz-mx} and in Figure~\ref{fig:periodogram-mxred} while we chose the model that has evidence closest to the median $\ln Z$ as our final model. The marginal evidence increases significantly, improving by 35.9 compared to the GP-only model, and by 18.4 and 19.8 relative to the 2-Planet models, depending on which two planets are modeled. The phase-folded plots for these three planets are shown in Figure~\ref{fig:phase-folded-mxred} in the Appendix. The periods and velocity semi-amplitudes ($K_{b}$ = $0.454\pm0.060$\,m\,s$^{-1}$, $K_{d}$ = $0.479\pm0.058$\,m\,s$^{-1}$, and $K_{c}$ = $0.467\pm0.058$\,m\,s$^{-1}$) derived from the analysis of the MAROON-X Red arm data for all three signals are consistent with the results of \citet{gonzalez24} within 1.5$\sigma$. Therefore, we confirm planet b and planet candidates c and d independently using MAROON-X alone.

Additionally, although the periodogram of the 3-Planet model does not show any significant signal, it shows some power around the period of planet candidate e ($P$\,=\,6.739\,d). A four-planet fit centered at this period provides a comparable value for the marginal evidence with the solution converging to same period as identified by \citet{gonzalez24}.

%The corner plot for the 3-planet model is presented in Figure~\ref{fig:corner-mxr} in Appendix~\ref{app:corner}. We note that there are a few injections at 3.177 d, the 1-year alias of the 3.15 d signal, possibly due to degeneracies between 38 different parameters in our 3-planet model. However, as these injections are $<1\%$ of total injections and the fact that we do not see this in a joint fit between MAROON-X Red channel and ESPRESSO data, we believe that 3.15 d signal is the true signal. We also acknowledge that the semi-amplitudes of all three planets are different from the values determined in a joint 4-planet fit between MAROON-X Red and ESPRESSO data. We attribute this to the fourth planet at 6.74 d period that is not modeled here. 

%We repeat the process using both Red and Blue channel data from MAROON-X. Although the Blue channel data doesn't show clear signals at the locations of these three planets, possibly due to reasons discussed in \S~\ref{subsec:stellar-activity}, we are able to recover accurate parameters by performing a joint fit with the Red channel data. The phase-folded plots of the three planets from the joint fit with both Blue and Red channel data along with the periodograms are presented in Figure~\ref{fig:phase-folded-rb}, \ref{fig:periodogram-rb} in Appendix~\ref{app:MX-rb}.

%Basant et al. (submitted) present an alternative calibration of the MAROON-X velocity zero point

\subsection{Joint analysis of MAROON-X and ESPRESSO}
Given the consistency of our results compared to \citet{gonzalez24}, we then turned our attention to joint modeling of the MAROON-X and ESPRESSO data to further constrain the system. We adopted the 149 radial velocities (which were derived via the S-BART code, \citealt{Silva_2022}) of the ESPRESSO data from \citet{gonzalez24}. Joint periodograms of the MAROON-X Red arm ESPRESSO radial velocities with successive planet signals removed are shown in Figure~\ref{fig:periodogram-combined}. The evidence for each model is given in Table~\ref{tab:lnz-combined}.

We ran the 4-Keplerian model ten times to analyze any variance in marginal evidence. The $\ln Z$ ranges between [$-328.7$, $-320.3$] with a median and standard-deviation of $-325.6\pm2.6$. The posteriors are consistent within different runs. However, For the 4-Keplerian model, the runs with the lowest $\ln Z$ have a slightly larger scatter ($< 0.03\%$ of the samples) outside the period range 6.72 to 6.76 in comparison to the fit with the highest marginal evidence ($< 0.005\%$ of the samples). We identify this as the major source of the observed variance in $\ln Z$. We report the median values of $\ln Z$ in Table~\ref{tab:lnz-combined} and in Figure~\ref{fig:periodogram-combined}, and we select the model with evidence closest to the median $\ln Z$ value as our final model.

%\textbf{We ran GP-only, 1-Keplerian, and 2-Keplerian models three times, the 3-Keplerian model five times, and the 4-Keplerian model ten times to analyze any variance in marginal evidence. We report the median values of lnZ in Table~\ref{tab:lnz-combined} and in Figure~\ref{fig:periodogram-combined} while we select the model with evidence closest to the median lnZ value as our final model. The ranges along with median and standard deviation for lnZ of all models are as follows: [$-420.0$, $-419.7$] and $-419.8\pm0.1$ for the GP-only model, [$-393.1$, $-392.9$] and $-393.0\pm0.1$ for the 1-Keplerian model, [$-371.8$, $-370.3$] and $-370.3\pm0.7$ for the 2-Keplerian model, [$-336.1$, $-331.2$] and $-332.8\pm1.8$ for the 3-Keplerian model, and [$-328.7$, $-320.3$] and $-325.6\pm2.6$ for the 4-Keplerian model. The posteriors are consistent within different runs of the same model. However, For the 4-Keplerian model, the runs with the lowest lnZ have a slightly larger scatter ($< 0.03\%$ of the samples) outside the period range 6.72 to 6.76 in comparison to the fit with the highest marginal evidence ($< 0.005\%$ of the samples). We identify this as the major source of the observed variance in lnZ. Similarly, for other models, we associate the variance observed in the marginal evidence to a few samples that are drawn around the 1-year aliases of planets b and c. The variance in marginal evidence was also reported by \cite{gonzalez24} for their eccentric 1-Keplerian model (E1e).}

The signals for planets b, c, and d show up clearly in the periodograms, and models with increasing planet numbers have very large improvements in the evidence. Furthermore, the signal for planet candidate e in the periodogram is strengthened compared to the ESPRESSO-only results (FAP = 1E-7\% vs.\ 2\%), and the change in evidence between the three- and four-planet models is highly significant ($\Delta\ln Z = 7.3$). We therefore also consider planet candidate e confirmed as a bona-fide planet.

The final parameters for the four-planet model are given in Table~\ref{tab:planets+GP-combined}, the phase-folded velocities are shown in Figure~\ref{fig:phase-folded-combined}, and the instrumental parameters can be found in Table~\ref{tab:instrumetal-param} in the Appendix. The corner plot with 20 planetary parameters and 6 GP hyperparameters is shown in Figure~\ref{fig:corner} in the Appendix. No further significant peaks are seen in the combined MAROON-X Red channel and ESPRESSO residuals after subtracting the four-planet model. We also repeated this analysis for a Global GP model where the stellar activity is modeled under a common covariance matrix. All parameters are in excellent agreement with the instrument-by-instrument GP model. The median values of $\ln Z$ for different global-GP models are presented in Table~\ref{tab:lnz-combined} while the realized GP for the 4-Keplerian model is shown in Figure~\ref{fig:GP} in the Appendix.

%\textbf{We also repeat the same analysis with a global GP model where the stellar activity is modeled under a common global covariance matrix. We ran different models multiple times. The median values and the standard deviations for lnZ are as follows: $-412.3\pm0.5$ for the GP-only model, $-386.6\pm0.6$ for the 1-Keplerian model, $-365.2\pm1.4$ for the 2-Keplerian model, $-333.7\pm1.0$ for the 3-Keplerian model, and $-326.4\pm2.7$ for the 4-Keplerian model. We associate the observed variance in marginal evidence to a few samples that are drawn around 1-year aliases of planets c, d, and e. The change in marginal evidence from a 3-Keplerian model to a 4-Keplerian model is $\Delta$lnZ $= 7.3$, consistent with our instrument-by-instrument GP model. The posteriors for all planets and GP hyperparameters are in excellent agreement with the instrument-by-instrument GP models. However, as the ESPRESSO and MAROON-X Red instruments cover different wavelength passbands, we elected to use the instrument-by-instrument GP as our final model. }

%\begin{figure}[]
%    \centering
%    \includegraphics[width=1\columnwidth,keepaspectratio]{MX-ESP-GP-residuals.pdf} % Adjust the size to fit within the page
%    \caption{Same as Figure~1 but for a joint fit between MAROON-X Red channel and ESPRESSO radial velocities.}
%    \label{fig:mxred-esp-residuals-0p}
%\end{figure}

\begin{figure}[]
    \centering
    \includegraphics[width=1\columnwidth, keepaspectratio]{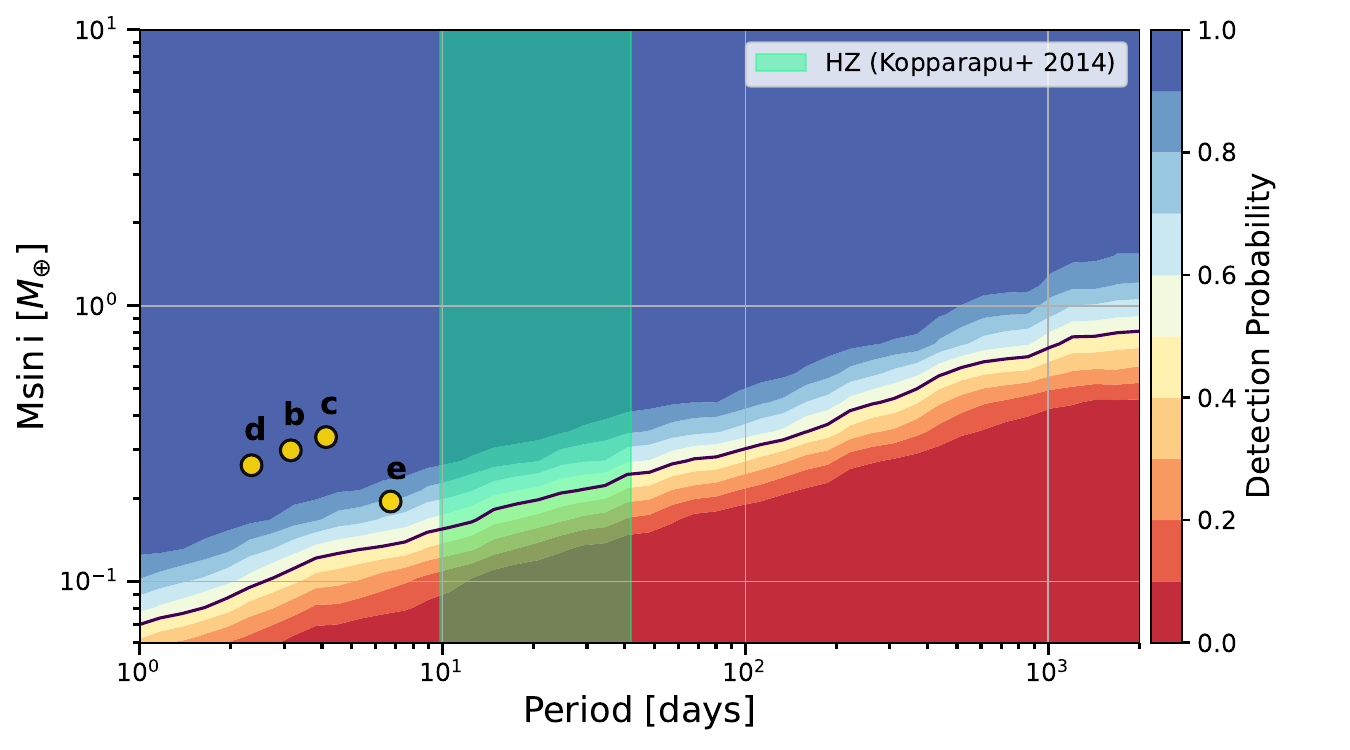} % Adjust the size to fit within the page
    \caption{This plot shows the current sensitivity of detecting planets around Barnard's Star using the MAROON-X Red channel and ESPRESSO data.}
    \label{fig:limits-HZ}
\end{figure}

\subsection{Stability analysis}\label{subsec:stability}
Our proposed four-planet configuration for Barnard's Star is remarkably compact (see discussion in \S4), thus raising the question of whether such a system would be stable. We investigated this by using the \texttt{SPOCK} tool \citep{tamayo2020} and numerical integrations using the \texttt{REBOUND} code \citep{rebound}. \texttt{SPOCK} is an algorithm that uses machine learning and resonant dynamics to assess the probability of dynamical stability of compact systems with 3--5 planets. 

We found that our best-fit point estimate for the planets' parameters results in an unstable system within 2,000\,years. When mapping the parameter space, we found that the small, but non-zero eccentricities of the planets in the best-fit solution are likely to blame for this instability. Using \texttt{SPOCK}, we see that the system has a $<$80\% chance of being stable if any one of the planets besides e has $e>0.02$ even when all the others have $e$=0.

Motivated by these results, we ran a joint MAROON-X+ESPRESSO fit with the eccentricities fixed to zero. The best-fit eccentricities in Table~\ref{tab:planets+GP-combined} are all consistent with zero within 1.5$\sigma$. The evidence is somewhat higher for the non-zero eccentricity fit ($\Delta\ln Z$ = 1.2), but there is a known bias to non-zero eccentricity in radial velocity fits \citep{laughlin05}.

We found that the four-planet system remains stable over 10\textsuperscript{9} orbits of the shortest-period planet ($P$=2.340\,d) when zero eccentricities are assumed for all the planets.  This holds true for planet masses up to three times the values given in Table~\ref{tab:planets+GP-combined}, meaning that the system is stable for planet inclinations ranging from 20 to 90 degrees.  Even in cases of low inclination, \texttt{SPOCK} returns an over 90\% probability that the system is stable. We therefore conclude that our proposed four-planet configuration is not obviously unstable and that eccentricities $<$0.02 are favored. Nevertheless, a more detailed analysis of the system's stability is warranted.

\begin{figure*}[]
    \centering
    \includegraphics[width=1\textwidth,height=1.0\textheight,keepaspectratio]{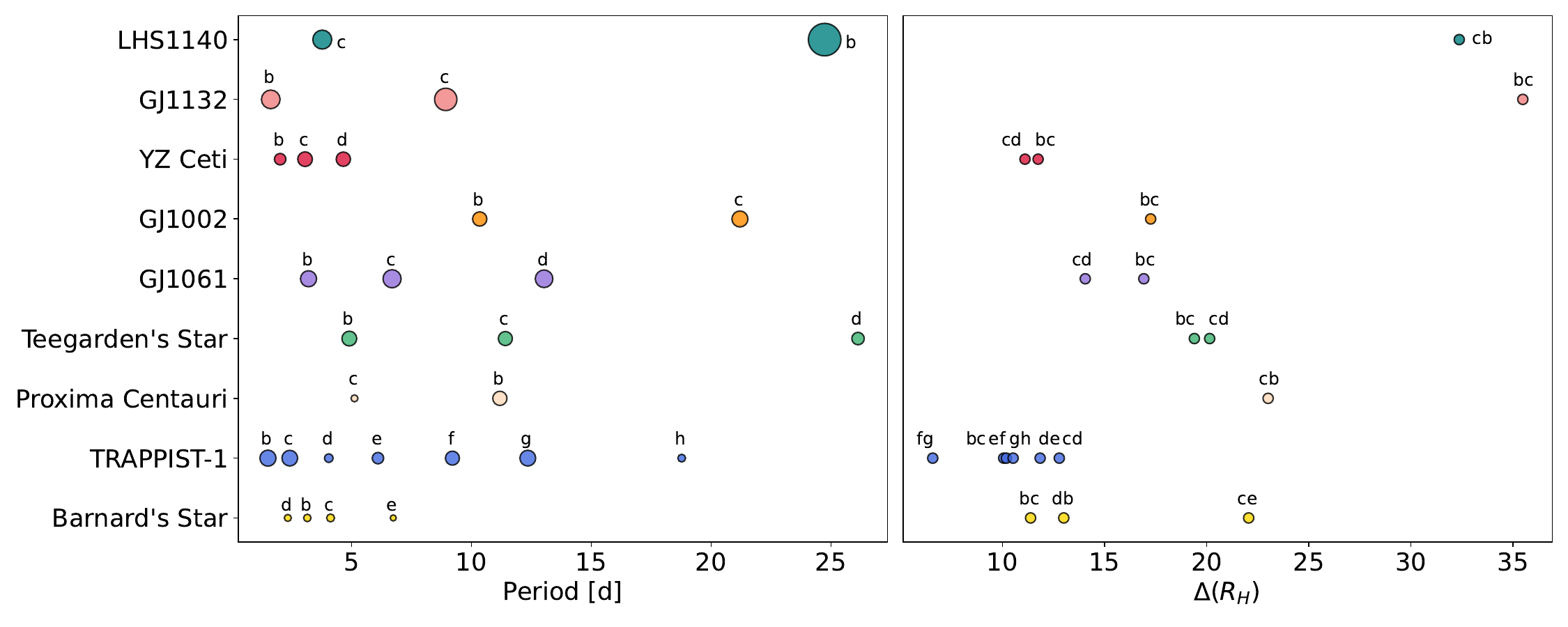} % Adjust the size to fit within the page
    \caption{This plot shows a comparison of planetary architectures between Barnard's Star and other compact M dwarf systems. Left Panel: Orbital periods of different planets with symbol sizes proportional to the planets' masses. Right Panel: Mutual Hill separations between consecutive planets in the systems.}
    \label{fig:mutual-hill}
\end{figure*}

\subsection{Limits on the habitable zone}
Based on the calculations of \citet{2013ApJ...765..131K,kopparapu14}, the habitable zone of Barnard's Star corresponds to orbital periods of 10 -- 42\,d assuming a 1\,M$_{\oplus}$ planet. The four detected planets are all closer to the star than the inner edge of the habitable zone, and there is no evidence of signals at longer periods. We therefore used the data to place limits on additional planets in the system.

We performed a suite of injection-recovery tests on the residuals of our 4-Planet joint fit of MAROON-X Red channel and ESPRESSO data using \texttt{rvsearch} \citep{2021ApJS..255....8R}. We injected 10,000 planetary signals where the semi-amplitudes, orbital periods, argument of periastron, and time of periastron passage were drawn from a uniform distribution while the eccentricities were drawn from a $\mathcal{\beta}$ distribution following \citet{2013MNRAS.434L..51K}. A planetary signal is classified as recovered if \texttt{rvsearch} recovered the orbital period and semi-amplitude within $25\%$ of the true values.

The results of our injection-recovery test are shown in Figure~\ref{fig:limits-HZ}. A similar analysis on the residuals of the 4-Planet fit to the ESPRESSO-only dataset reveals the detection probability of a planet e analogue (with parameters from Table~\ref{tab:planets+GP-combined}) to be 59\%. The detection probability of planet e in the residuals of the joint-fit between MAROON-X Red channel and ESPRESSO data improves to 79\%, thus making the detection of planet e more likely with the combined dataset. Based on our analysis, the current measurements allow us to rule out the presence of planets with minimum masses $0.37 M_{\oplus}$ to $0.57 M_{\oplus}$ (detection probability $> 0.99$) near the inner and outer edges of the habitable zone. Further observations are needed to investigate the presence of planets with even lower masses.  

\section{Discussion}\label{sec:discussion}
At long last Barnard's Star has planets confirmed with two different instruments. Demonstrating the power of this new generation of EPRV spectrographs, the planet signals are all $<$\,50\,cm\,s$^{-1}$, and the planets have minimum masses $<$\,0.34\,$M_{\oplus}$. Barnard e is in contention with Proxima d \citep{faria22} as the lowest-mass planet detected with radial velocities ($m\sin i = 0.19 \pm 0.03\,M_{\oplus}$ vs.\ $0.26 \pm 0.05\,M_{\oplus}$, respectively). With $K = 0.221 \pm 0.038$\,cm\,s$^{-1}$, Barnard e has the lowest radial velocity semi-amplitude of a claimed planet. We measure this signal to 6$\sigma$ confidence using MAROON-X and ESPRESSO.

The Barnard's Star system is remarkably compact. \citet{dreizler24} compared the values of mutual Hill radius separation ($\Delta(R_{H})$) of several multi-planetary systems hosted by late M-dwarfs that consisted of at least one potential rocky planet (< $2M_{\oplus}$). In this study, we conduct a similar analysis, incorporating Barnard's Star into the comparison. The left panel of Figure~\ref{fig:mutual-hill} compares the planets orbiting Barnard's Star to those in eight other compact multi-planet systems: (1) TRAPPIST-1 \citep{2021PSJ.....2....1A}, (2) GJ\,1132 \citep{2024ApJ...973L...8X, 2018A&A...618A.142B}, (3) YZ Ceti \citep{2020A&A...636A.119S}, (4) GJ\,1002 \citep{2023A&A...670A...5S}, (5) GJ\,1061 \citep{2020MNRAS.493..536D}, (6) Teegarden's Star \citep{dreizler24}, (7) Proxima Centauri \citep{faria22}, and (8) LHS\,1140 \citep{2024ApJ...960L...3C}. The marker sizes scale with the minimum mass (or actual mass for TRAPPIST-1 planets and GJ\,1132b). The right panel illustrates the mutual Hill radius separations for consecutive planets across all nine systems. The inner three planets orbiting Barnard's Star have comparable separations as the compact TRAPPIST-1 and YZ Ceti systems. As described in \S3.4, we find that the eccentricities of the Barnard's Star planets are all consistent with zero, thus likely aiding the dynamical stability of the system.

The radial velocity data for Barnard's Star show remarkably low scatter after just removing a simple model for rotationally-modulated activity. The MAROON-X Red channel residuals have an RMS of 31\,cm\,s$^{-1}$, which is essentially equivalent to their average photon-limited errors of 29\,cm\,s$^{-1}$. The ESPRESSO residuals have an RMS of 23\,cm\,s$^{-1}$, which is substantially larger than their typical uncertainties of 10\,cm\,s$^{-1}$. This could either be indicative of residual stellar activity or unidentified instrument systematics. Our experience with MAROON-X and our takeaways from the literature suggest that even the best EPRV instruments are still not achieving their maximum potential performance. For example, neither the MAROON-X data nor the ESPRESSO data were based on laser frequency comb calibration, and both instruments experience variations in temperature and pressure that could be controlled better. The detection of such small planets with large, but not overly intensive data sets obtained with sub-optimal instrument performance bodes well for the future of radial velocity planet searches. It is remarkable that radial velocities can still push the frontier of exoplanet detection even 30 years after the detection of 51 Peg b \citep{mayor95}.

\section*{Acknowledgements}
The University of Chicago group acknowledges funding for the MAROON-X project from the David and Lucile Packard Foundation, the Heising-Simons Foundation, the Gordon and Betty Moore Foundation, the Gemini Observatory, the NSF (award number 2108465), and NASA (grant number 80NSSC22K0117). RL and LLZ acknowledge support from NASA through the NASA Hubble Fellowship (grants HST-HF2-51559 and HST-HF2-51569) awarded by the Space Telescope Science Institute, which is operated by the Association of Universities for Research in Astronomy, Inc., for NASA, under contract NAS5-26555. The Gemini observations are associated with programs 21A-Q-119, 21A-Q-404, 21B-LP-202, 22A-CAL-201, 22A-LP-202, 22A-Q-409, 22B-CAL-201, 22B-Q-409, 23A-Q-120, 23A-Q-405, 23B-LP-202.

\vspace{5mm}
\facilities{Gemini-N (MAROON-X), VLT (ESPRESSO)}

\bibliography{ms}{}
\bibliographystyle{aasjournal}

\appendix
\begin{longtable}{ccccc}
\caption{MAROON-X RV dataset of Barnard's Star} \label{table:Barnards-MX-rvs} \\
\hline
\hline
BJD & RV MX Red & Error MX Red & RV MX Blue & Error MX Blue \\
\text{[d]} & \text{[\,m\,s$^{-1}$]} & \text{[\,m\,s$^{-1}$]} & \text{[\,m\,s$^{-1}$]} & \text{[\,m\,s$^{-1}$]} \\
\hline
\endfirsthead
\multicolumn{5}{c}%
{{\tablename\ \thetable{} -- continued from previous page}} \\
\hline
\hline
BJD & RV MX Red & Error MX Red & RV MX Blue & Error MX Blue \\
\text{[d]} & \text{[\,m\,s$^{-1}$]} & \text{[\,m\,s$^{-1}$]} & \text{[\,m\,s$^{-1}$]} & \text{[\,m\,s$^{-1}$]} \\
\hline
\endhead

\hline
\multicolumn{5}{r}{{Continued on next page}} \\
\endfoot

\hline
\multicolumn{5}{r}{{End of table}} \\
\endlastfoot

2459321.089339 & -8.48 & 0.32 & -7.64 & 0.42 \\
2459322.110600 & -9.66 & 0.34 & -8.33 & 0.46 \\
2459324.112461 & -10.01 & 0.29 & -9.16 & 0.37 \\
2459325.114176 & -8.75 & 0.30 & -8.45 & 0.38 \\
2459327.100589 & -9.50 & 0.32 & -8.04 & 0.41 \\
2459328.109813 & -8.67 & 0.36 & -8.03 & 0.50 \\
2459333.007912 & -8.23 & 0.32 & -5.83 & 0.43 \\
2459333.969222 & -7.70 & 0.30 & -6.44 & 0.40 \\
2459335.009764 & -7.07 & 0.28 & -5.93 & 0.35 \\
2459358.965661 & -5.06 & 0.33 & -4.74 & 0.50 \\
2459359.919573 & -6.34 & 0.34 & -4.71 & 0.52 \\
2459360.121373 & -5.67 & 0.19 & -4.30 & 0.29 \\
2459361.030219 & -6.50 & 0.34 & -4.18 & 0.51 \\
2459361.948153 & -7.03 & 0.28 & -5.38 & 0.39 \\
2459362.979858 & -6.01 & 0.25 & -4.83 & 0.31 \\
2459363.944908 & -7.38 & 0.26 & -5.46 & 0.36 \\
2459365.032154 & -7.05 & 0.24 & -6.18 & 0.31 \\
2459366.047448 & -6.26 & 0.19 & -4.84 & 0.26 \\
2459367.114880 & -6.87 & 0.26 & -5.17 & 0.40 \\
2459368.008443 & -5.47 & 0.27 & -4.09 & 0.36 \\
2459368.907788 & -6.57 & 0.25 & -5.42 & 0.35 \\
2459440.742736 & -4.98 & 0.30 & -3.94 & 0.38 \\
2459441.799269 & -5.09 & 0.28 & -3.48 & 0.33 \\
2459444.906483 & -4.60 & 0.29 & -3.60 & 0.41 \\
2459446.781749 & -5.08 & 0.34 & -4.04 & 0.45 \\
2459447.847732 & -4.05 & 0.39 & -2.76 & 0.58 \\
2459448.775678 & -5.15 & 0.22 & -2.98 & 0.34 \\
2459449.826562 & -4.25 & 0.27 & -3.59 & 0.32 \\
2459514.722096 & -1.70 & 0.31 & 0.20 & 0.46 \\
2459524.690827 & -1.19 & 0.37 & 1.10 & 0.57 \\
2459663.045448 & 0.97 & 0.29 & 2.96 & 0.47 \\
2459664.139126 & 0.89 & 0.27 & -0.90 & 0.39 \\
2459665.034895 & 2.31 & 0.22 & 0.26 & 0.28 \\
2459666.144693 & 0.04 & 0.25 & 1.24 & 0.32 \\
2459667.130997 & 0.42 & 0.28 & 1.75 & 0.39 \\
2459671.144615 & 0.84 & 0.27 & 2.73 & 0.36 \\
2459672.030593 & 2.38 & 0.19 & 1.30 & 0.26 \\
2459673.141590 & 1.60 & 0.27 & 2.47 & 0.36 \\
2459674.020896 & 2.12 & 0.21 & 3.21 & 0.30 \\
2459674.148511 & 2.52 & 0.38 & 3.43 & 0.57 \\
2459678.001730 & 2.19 & 0.30 & 3.35 & 0.40 \\
2459678.094617 & 2.18 & 0.29 & 4.68 & 0.39 \\
2459679.038478 & 1.58 & 0.30 & 3.47 & 0.41 \\
2459679.139257 & 1.96 & 0.31 & 4.70 & 0.43 \\
2459681.016456 & 2.85 & 0.18 & 5.21 & 0.27 \\
2459681.139209 & 3.17 & 0.35 & 5.27 & 0.51 \\
2459682.034007 & 2.38 & 0.40 & 4.52 & 0.62 \\
2459682.981733 & 0.87 & 0.30 & 2.91 & 0.40 \\
2459683.106694 & 1.62 & 0.33 & 2.56 & 0.45 \\
2459684.140664 & 3.03 & 0.29 & 5.72 & 0.36 \\
2459685.135344 & 1.66 & 0.29 & 3.32 & 0.36 \\
2459687.984864 & 1.47 & 0.19 & 2.98 & 0.29 \\
2459689.029985 & 1.05 & 0.32 & 2.72 & 0.44 \\
2459690.124045 & 1.29 & 0.28 & 3.12 & 0.33 \\
2459691.105625 & 1.09 & 0.29 & 3.22 & 0.34 \\
2459693.070956 & 2.22 & 0.32 & 3.99 & 0.41 \\
2459693.950918 & 1.56 & 0.25 & 3.51 & 0.37 \\
2459695.097745 & 0.15 & 0.29 & 1.88 & 0.35 \\
2459696.065464 & 0.52 & 0.29 & 2.00 & 0.35 \\
2459697.014077 & 1.62 & 0.29 & 2.73 & 0.35 \\
2459724.013139 & 3.87 & 0.25 & 2.30 & 0.33 \\
2459725.131552 & 4.41 & 0.27 & 3.15 & 0.37 \\
2459726.013637 & 5.31 & 0.25 & 3.42 & 0.33 \\
2459726.937287 & 3.86 & 0.24 & 1.76 & 0.31 \\
2459727.934069 & 4.72 & 0.32 & 2.87 & 0.44 \\
2459728.100608 & 4.96 & 0.17 & 3.24 & 0.23 \\
2459729.935123 & 3.18 & 0.26 & 1.53 & 0.32 \\
2459730.924961 & 5.78 & 0.26 & 3.38 & 0.34 \\
2459731.896371 & 4.32 & 0.25 & 2.34 & 0.33 \\
2459733.036065 & 4.41 & 0.18 & 2.87 & 0.24 \\
2459768.885438 & 1.56 & 0.30 & 0.26 & 0.40 \\
2459768.982170 & 1.84 & 0.34 & -0.30 & 0.49 \\
2459769.974648 & 2.78 & 0.36 & 0.24 & 0.53 \\
2459770.900837 & 2.23 & 0.28 & -0.64 & 0.36 \\
2459771.868236 & 2.37 & 0.25 & -0.00 & 0.30 \\
2459772.000134 & 2.68 & 0.29 & -0.01 & 0.37 \\
2459772.922860 & 2.56 & 0.29 & 0.86 & 0.38 \\
2459774.016654 & 0.56 & 0.32 & -0.39 & 0.50 \\
2459774.972000 & 1.25 & 0.36 & 0.24 & 0.50 \\
2459775.811500 & 1.53 & 0.27 & -1.82 & 0.33 \\
2459778.846142 & 0.95 & 0.32 & -1.93 & 0.42 \\
2459779.979551 & 1.26 & 0.31 & -1.56 & 0.40 \\
2459780.916053 & -0.18 & 0.29 & -3.44 & 0.36 \\
2459781.977301 & 0.19 & 0.30 & -2.71 & 0.38 \\
2459786.869912 & 0.56 & 0.28 & -2.63 & 0.34 \\
2459787.762706 & 0.24 & 0.28 & -3.73 & 0.34 \\
2459791.955902 & 1.73 & 0.35 & -2.12 & 0.47 \\
2459800.964511 & 4.20 & 0.26 & 0.56 & 0.41 \\
2459801.819144 & 1.65 & 0.21 & -1.34 & 0.31 \\
2460116.920705 & 5.74 & 0.19 & 4.25 & 0.27 \\
2460117.873013 & 5.33 & 0.28 & 3.11 & 0.37 \\
2460118.934507 & 5.61 & 0.26 & 4.14 & 0.31 \\
2460119.787149 & 5.11 & 0.25 & 3.49 & 0.31 \\
2460120.882366 & 5.12 & 0.27 & 3.79 & 0.35 \\
2460121.891302 & 5.66 & 0.30 & 4.27 & 0.42 \\
2460123.001395 & 6.45 & 0.34 & 5.27 & 0.49 \\
2460124.007448 & 3.67 & 0.26 & 2.54 & 0.34 \\
2460124.931171 & 3.65 & 0.19 & 2.36 & 0.25 \\
2460125.945157 & 5.42 & 0.38 & 4.44 & 0.56 \\
2460127.982584 & 3.29 & 0.27 & 2.07 & 0.33 \\
2460129.017544 & 4.03 & 0.36 & 3.87 & 0.51 \\
2460129.908421 & 3.88 & 0.16 & 1.87 & 0.22 \\
2460130.833025 & 4.41 & 0.30 & 1.67 & 0.39 \\
2460131.845693 & 2.22 & 0.26 & 0.57 & 0.32 \\
2460133.951563 & 1.75 & 0.26 & 0.12 & 0.31 \\
2460134.864425 & 4.09 & 0.20 & 2.01 & 0.28 \\
2460135.825284 & 3.53 & 0.22 & 1.85 & 0.31 \\
2460136.794707 & 1.22 & 0.30 & -0.32 & 0.39 \\
2460231.740856 & 12.38 & 0.37 & 9.49 & 0.47 \\
2460240.753485 & 13.15 & 0.39 & 10.64 & 0.53 \\
2460242.699704 & 14.68 & 0.40 & 11.10 & 0.54 \\
2460243.718891 & 13.09 & 0.40 & 10.14 & 0.55 \\
\label{table:dataset}
\end{longtable}

\begin{table}[]
    \centering
    \begin{tabular}{ccc}
    \hline
    \hline
    \multicolumn{3}{c}{Instrumental Parameters}\\
    \hline
    Parameter & Posterior & Prior Distribution \\
    \hline
        $\gamma_{\rm ESP18}$ & $0.22_{-1.45}^{+1.44}$ & $\mathcal{U}$ $[-20, 20]$\\
        $\gamma_{\rm ESP19}$ & $0.19_{-0.37}^{+0.36}$ & $\mathcal{U}$ $[-20, 20]$\\
        $\gamma_{\rm 2021Apr-MXRed}$ & $-8.53_{-1.65}^{+1.64}$ & $\mathcal{U}$ $[-20, 20]$\\
        $\gamma_{\rm 2021May-MXRed}$ & $-5.63_{-1.65}^{+1.70}$ & $\mathcal{U}$ $[-20, 20]$\\
        $\gamma_{\rm 2021Aug-MXRed}$ & $-5.22_{-1.72}^{+1.69}$ & $\mathcal{U}$ $[-20, 20]$\\
        $\gamma_{\rm 2021Nov-MXRed}$ & $-1.66_{-1.86}^{+1.85}$ & $\mathcal{U}$ $[-20, 20]$\\
        $\gamma_{\rm 2022Apr-MXRed}$ & $1.00_{-1.32}^{+1.31}$ & $\mathcal{U}$ $[-20, 20]$\\
        $\gamma_{\rm 2022May-MXRed}$ & $4.45_{-1.73}^{+1.69}$ & $\mathcal{U}$ $[-20, 20]$\\
        $\gamma_{\rm 2022Jul-MXRed}$ & $2.62_{-1.33}^{+1.37}$ & $\mathcal{U}$ $[-20, 20]$\\
        $\gamma_{\rm 2023Jul-MXRed}$ & $3.89_{-1.56}^{+1.52}$ & $\mathcal{U}$ $[-20, 20]$\\
        $\gamma_{\rm 2023Oct-MXRed}$ & $13.35_{-1.67}^{+1.71}$ & $\mathcal{U}$ $[-20, 20]$\\
        $\sigma_{\rm ESP18}$ & $0.30_{-0.13}^{+0.21}$ & $\mathcal{LU}$ $[0.1, 5.0]$\\
        $\sigma_{\rm ESP19}$ & $0.30_{-0.03}^{+0.03}$ & $\mathcal{LU}$ $[0.1, 5.0]$\\
        $\sigma_{\rm 2021Apr-MXRed}$ & $0.19_{-0.07}^{+0.14}$ & $\mathcal{LU}$ $[0.1, 5.0]$\\
        $\sigma_{\rm 2021May-MXRed}$ & $0.19_{-0.07}^{+0.12}$ & $\mathcal{LU}$ $[0.1, 5.0]$\\
        $\sigma_{\rm 2021Aug-MXRed}$ & $0.18_{-0.06}^{+0.14}$ & $\mathcal{LU}$ $[0.1, 5.0]$\\
        $\sigma_{\rm 2021Nov-MXRed}$ & $0.39_{-0.24}^{+0.91}$ & $\mathcal{LU}$ $[0.1, 5.0]$\\
        $\sigma_{\rm 2022Apr-MXRed}$ & $0.16_{-0.04}^{+0.07}$ & $\mathcal{LU}$ $[0.1, 5.0]$\\
        $\sigma_{\rm 2022May-MXRed}$ & $0.41_{-0.13}^{+0.18}$ & $\mathcal{LU}$ $[0.1, 5.0]$\\
        $\sigma_{\rm 2022Jul-MXRed}$ & $0.17_{-0.05}^{+0.09}$ & $\mathcal{LU}$ $[0.1, 5.0]$\\
        $\sigma_{\rm 2023Jul-MXRed}$ & $0.38_{-0.10}^{+0.13}$ & $\mathcal{LU}$ $[0.1, 5.0]$\\
        $\sigma_{\rm 2023Oct-MXRed}$ & $0.23_{-0.10}^{+0.27}$ & $\mathcal{LU}$ $[0.1, 5.0]$\\
        \hline
    \end{tabular}
    \caption{This table summarizes the posteriors and priors for the instrumental parameters for the final 4-Keplerian joint fit between MAROON-X and ESPRESSO datasets.}
    \label{tab:instrumetal-param}
\end{table}

\begin{figure*}[]
    \centering
    \includegraphics[width=0.7\columnwidth, keepaspectratio]{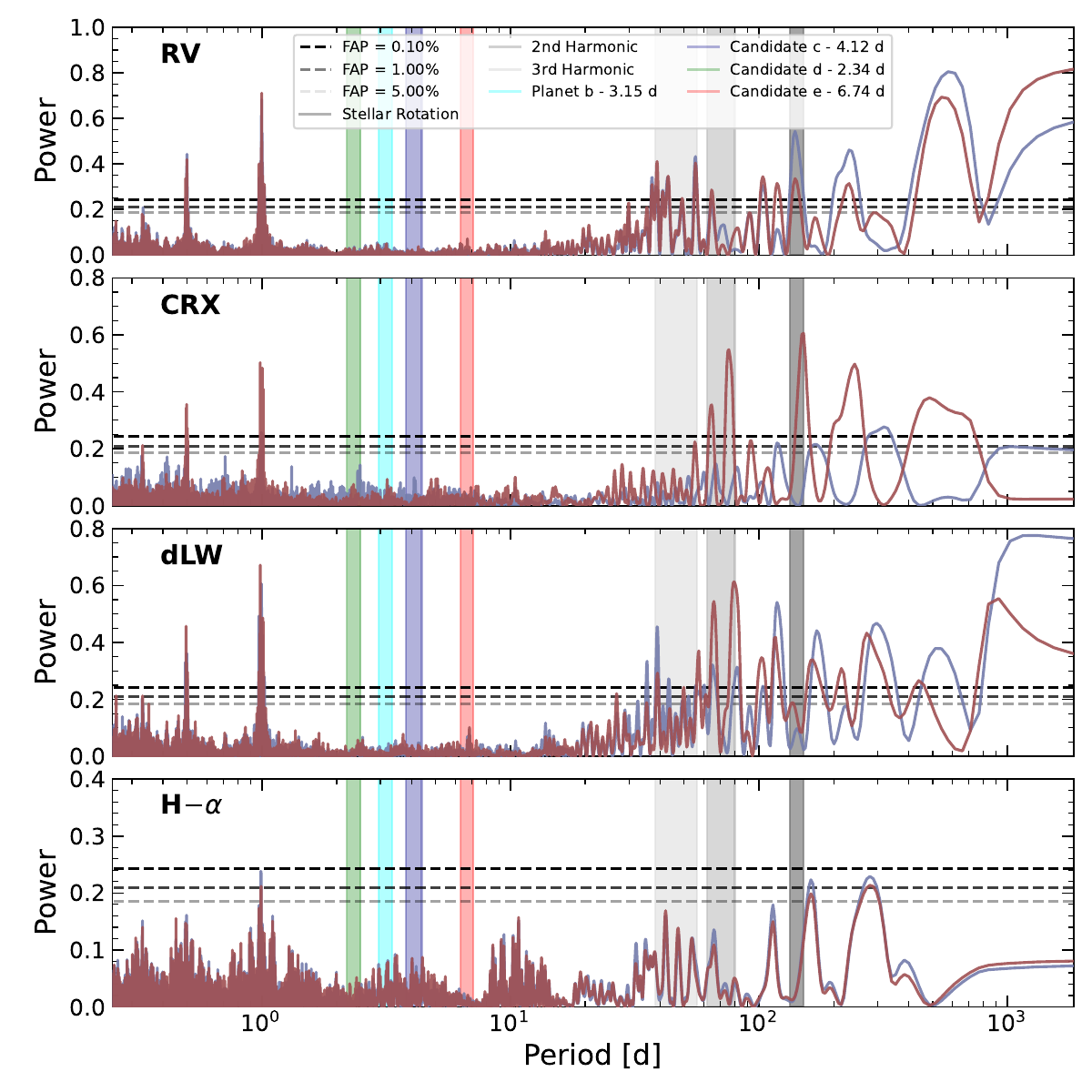} % Adjust the size to fit within the page
    \caption{This plot shows the computed GLS periodograms of raw radial velocities (red color for Red channel data and blue color for Blue channel data, chromatic index (CRX), differential line widths (dLW), and Halpha index for MAROON-X Red and Blue channel data. }
    \label{fig:activity}
\end{figure*}

\begin{figure*}[]
    \centering
    \includegraphics[width=1\columnwidth, keepaspectratio]{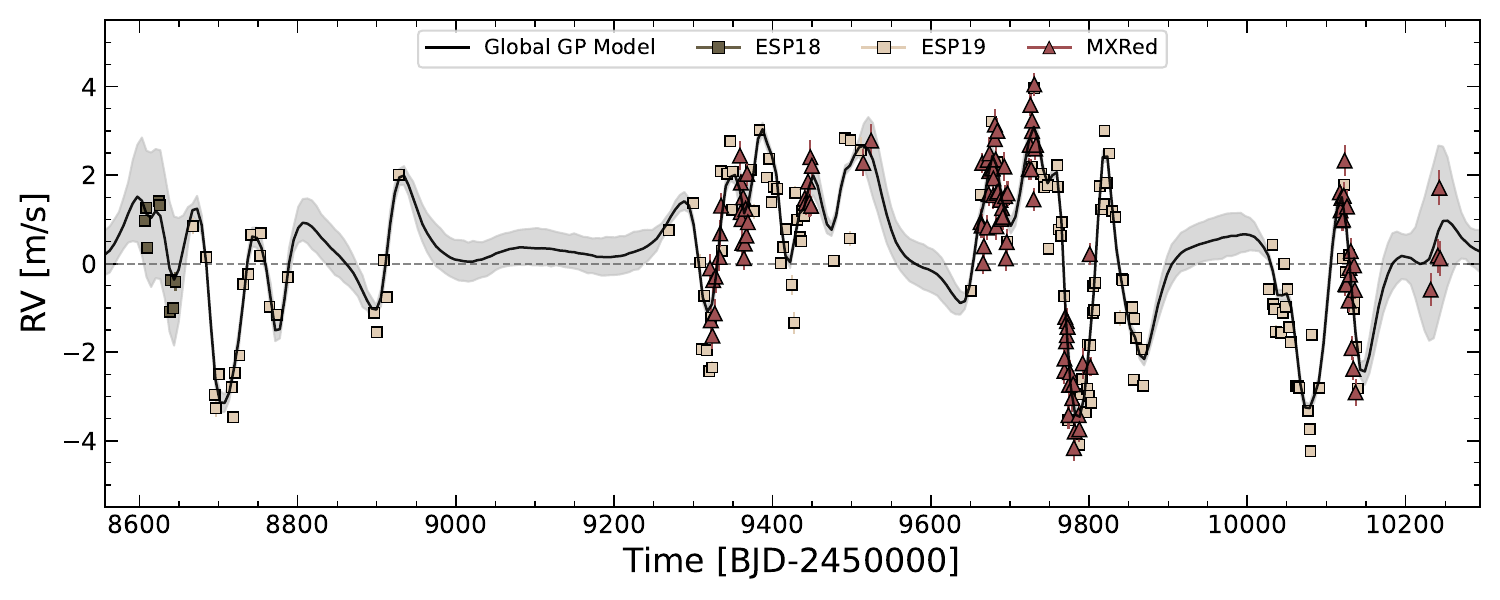} % Adjust the size to fit within the page
    \caption{This plot shows the Global GP model for the 4-Keplerian model joint fit between the MAROON-X Red channel and ESPRESSO datasets.}
    \label{fig:GP}
\end{figure*}

\begin{figure*}[]
    \centering
    \includegraphics[width=0.9\textwidth,height=0.8\textheight,keepaspectratio]{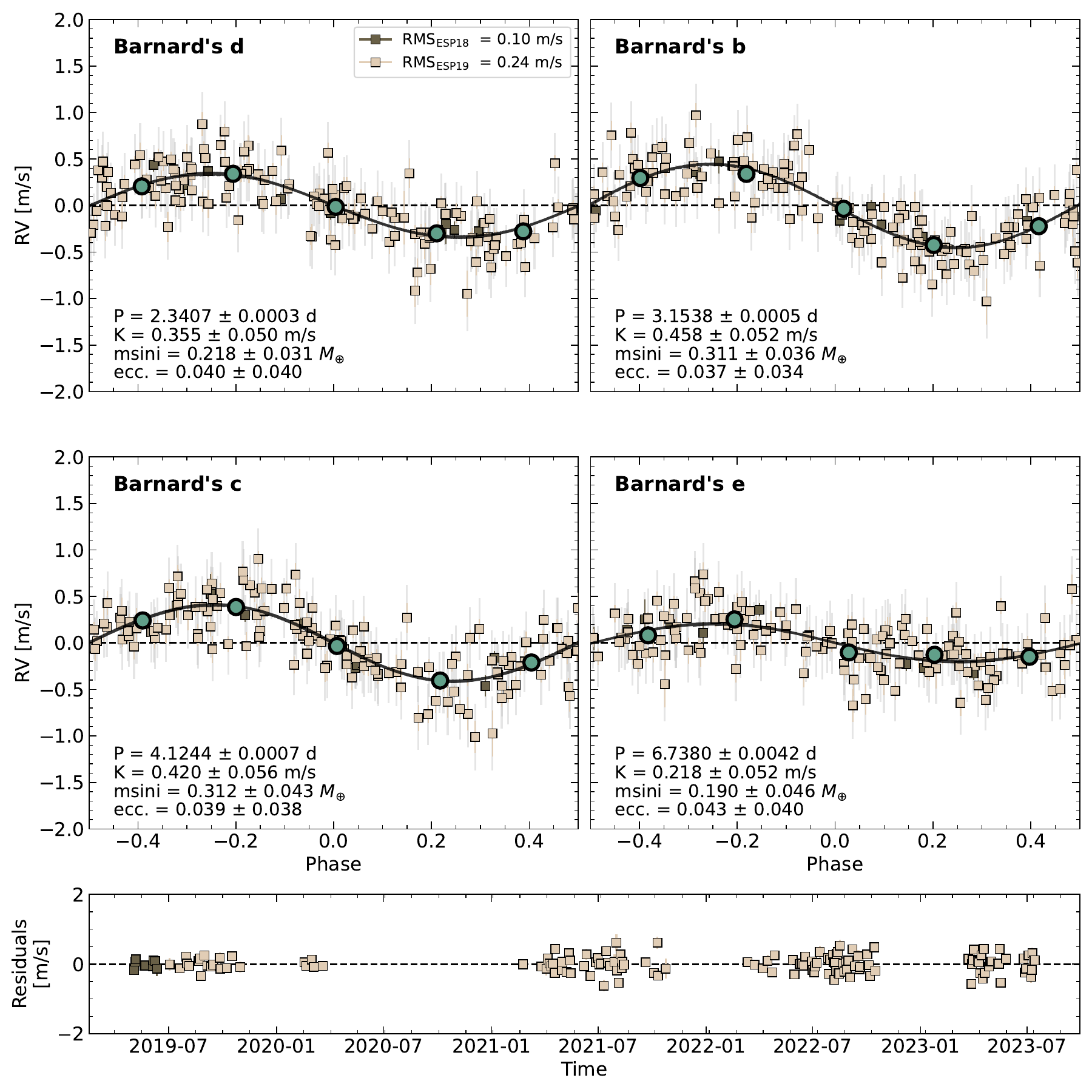} % Adjust the size to fit within the page
    \caption{Phase-folded plots for planets Barnard b, c, d, and e using ESPRESSO data.}
    \label{fig:phase-folded-esp}
\end{figure*}

\begin{figure*}[ht!]
    \centering
    \includegraphics[width=1.0\textwidth,height=1.0\textheight,keepaspectratio]{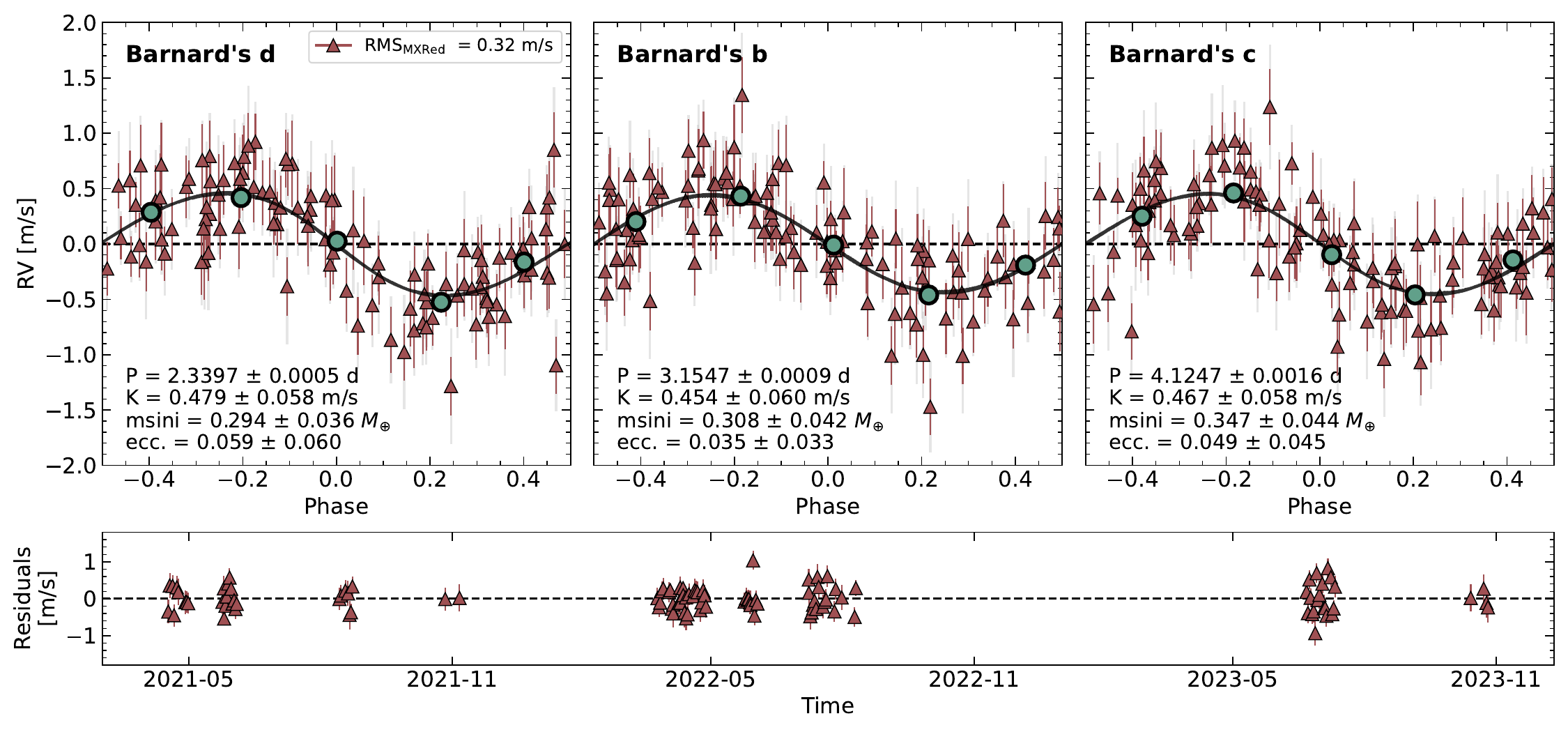} % Adjust the size to fit within the page
    \caption{Phase-folded plots for planets Barnard b, c, and d based on MAROON-X Red Channel data. The bottom panel shows the residuals of the 3-Planet model for MAROON-X Red channel radial velocities.}
    \label{fig:phase-folded-mxred}
\end{figure*}

\begin{figure*}[]
    \centering
    \includegraphics[width=1\textwidth,height=0.8\textheight,keepaspectratio]{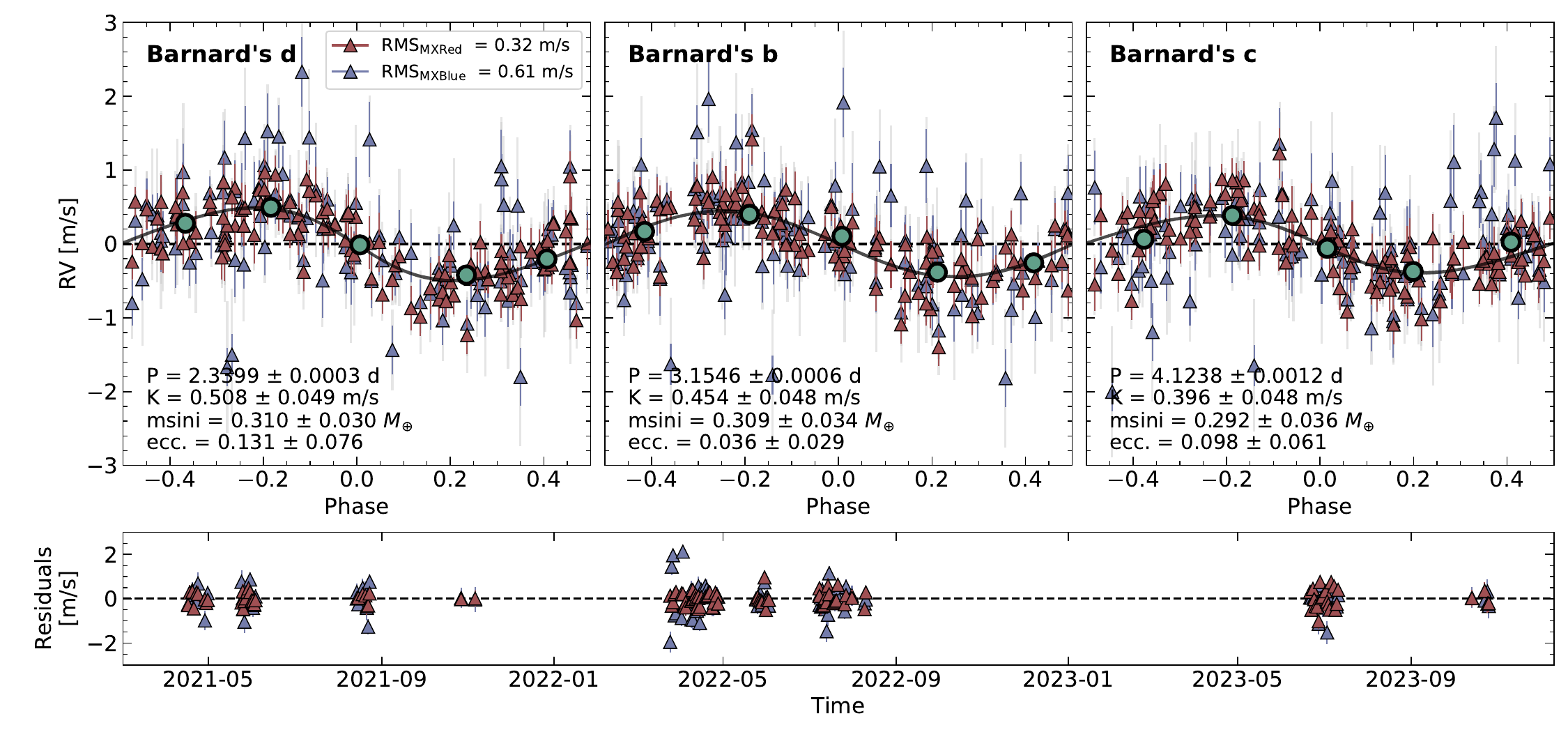} % Adjust the size to fit within the page
    \caption{Phase-folded plots for planets Barnard b, c, d, and e based on joint fit between MAROON-X Red and Blue channel radial velocities.}
    \label{fig:phase-folded-rb}
\end{figure*}

\begin{figure}[]
    \centering
    \includegraphics[width=0.7\textwidth, keepaspectratio]{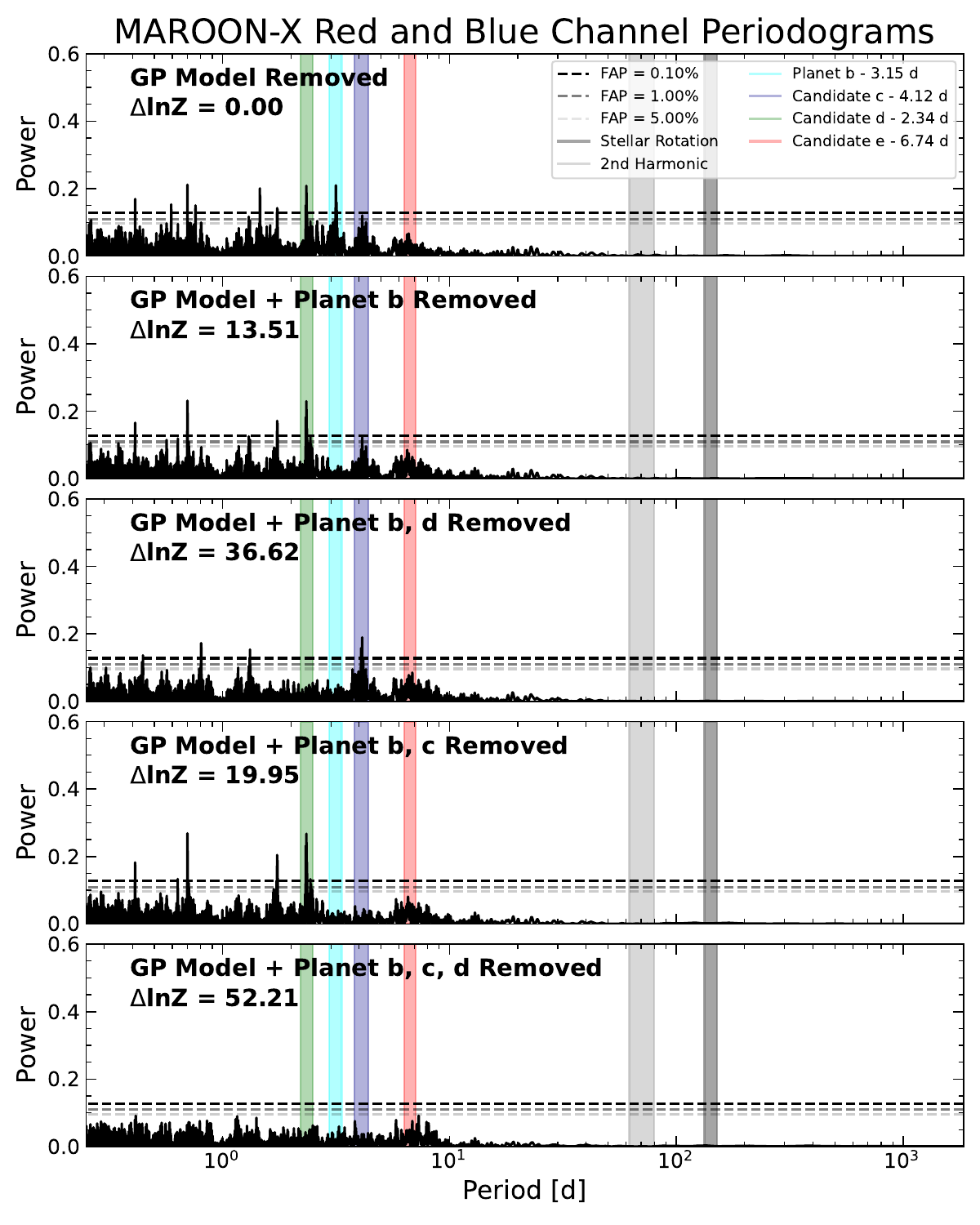} % Adjust the size to fit within the page
    \caption{This plot shows the computed GLS periodograms for the combined residuals of MAROON-X Red and Blue channel radial velocities. A beta distribution for the planetary eccentricities has been used for all planets. }
    \label{fig:periodogram-rb}
\end{figure}

%\begin{figure*}[]
%    \centering
%    \includegraphics[width=0.8\textwidth,height=0.8\textheight,keepaspectratio]{keplerian_corner_plots-mxred.pdf} % Adjust the size to fit within the page
%    \caption{Corner plot for 3-planet fit to MAROON-X Red channel radial velocities.}
%    \label{fig:corner-mxr}
%\end{figure*}

\begin{figure}[]
    \centering
    \includegraphics[width=0.7\columnwidth, keepaspectratio]{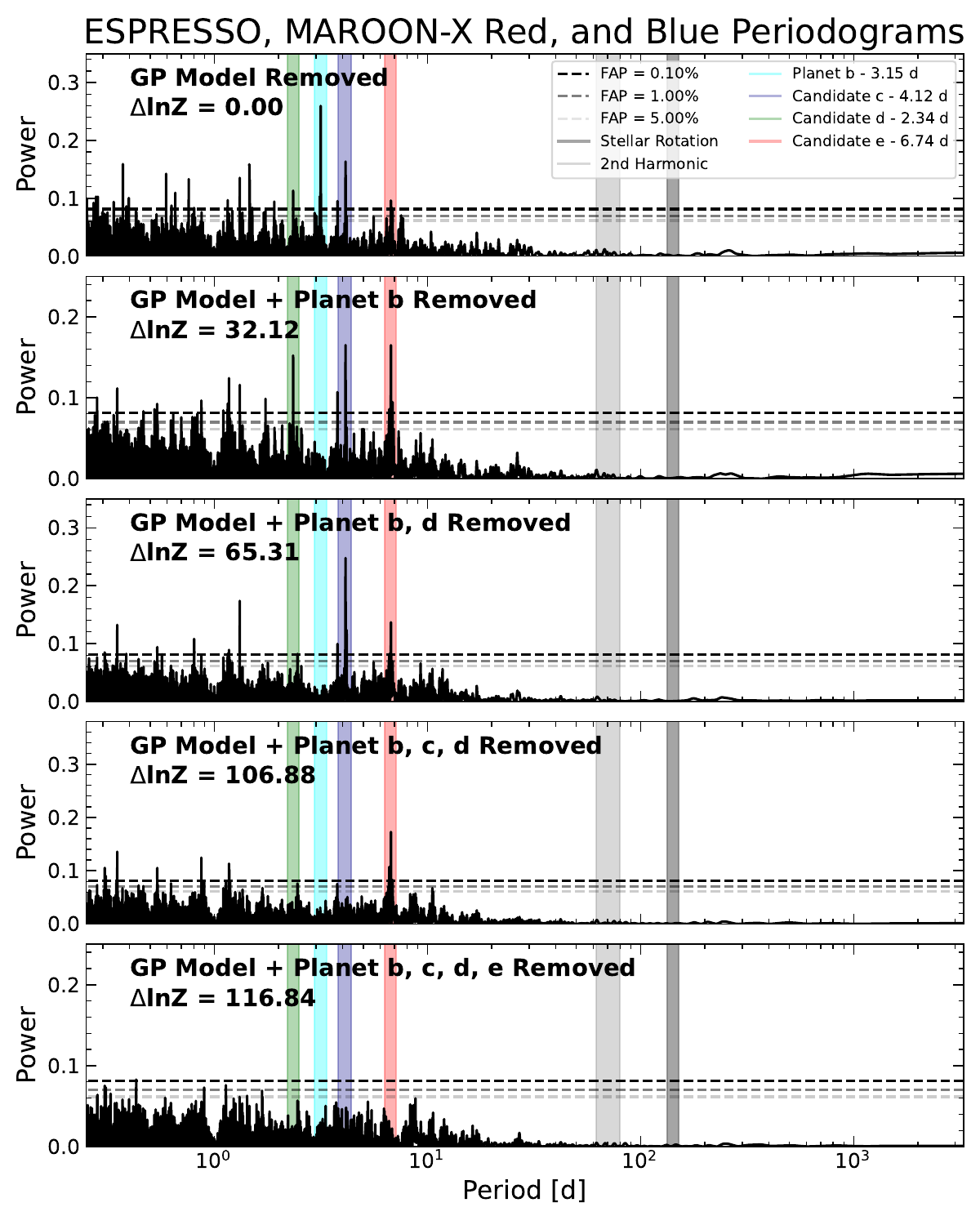} % Adjust the size to fit within the page
    \caption{This plot shows the computed GLS periodograms for the combined residuals of MAROON-X Red channel, Blue channel, and ESPRESSO radial velocities. A beta distribution for the planetary eccentricities has been used for all planets. }
    \label{fig:periodogram-mxrb-esp}
\end{figure}

\begin{figure*}[]
    \centering
    \includegraphics[width=0.9\textwidth,height=0.8\textheight,keepaspectratio]{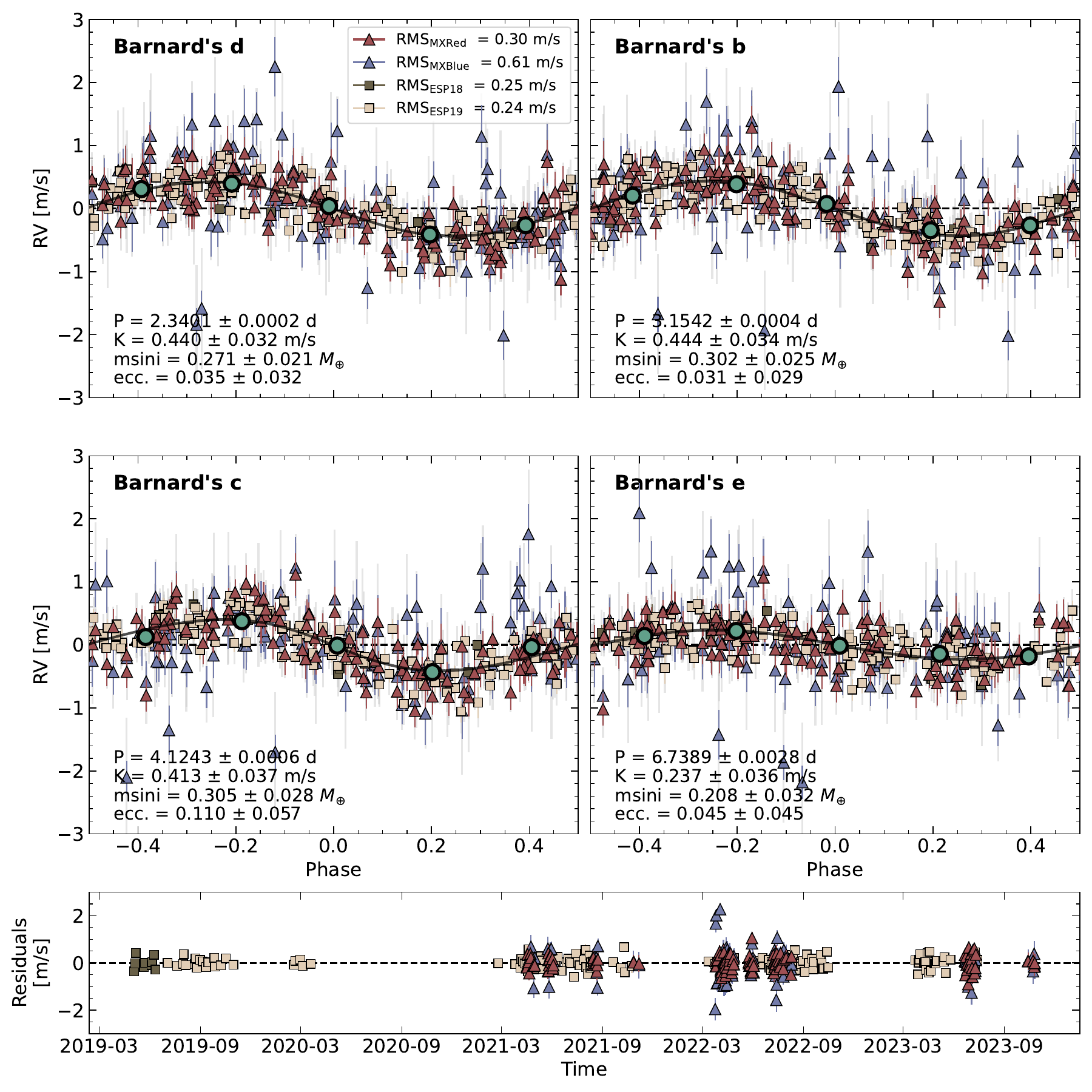} % Adjust the size to fit within the page
    \caption{Phase-folded plots for planets Barnard b, c, d, and e based on joint fit between MAROON-X Red channel, Blue channel, and ESPRESSO radial velocities.}
    \label{fig:phase-folded-mxrb-esp}
\end{figure*}

\begin{figure*}[]
    \centering
\includegraphics[width=1\textwidth,height=1.0\textheight,keepaspectratio]{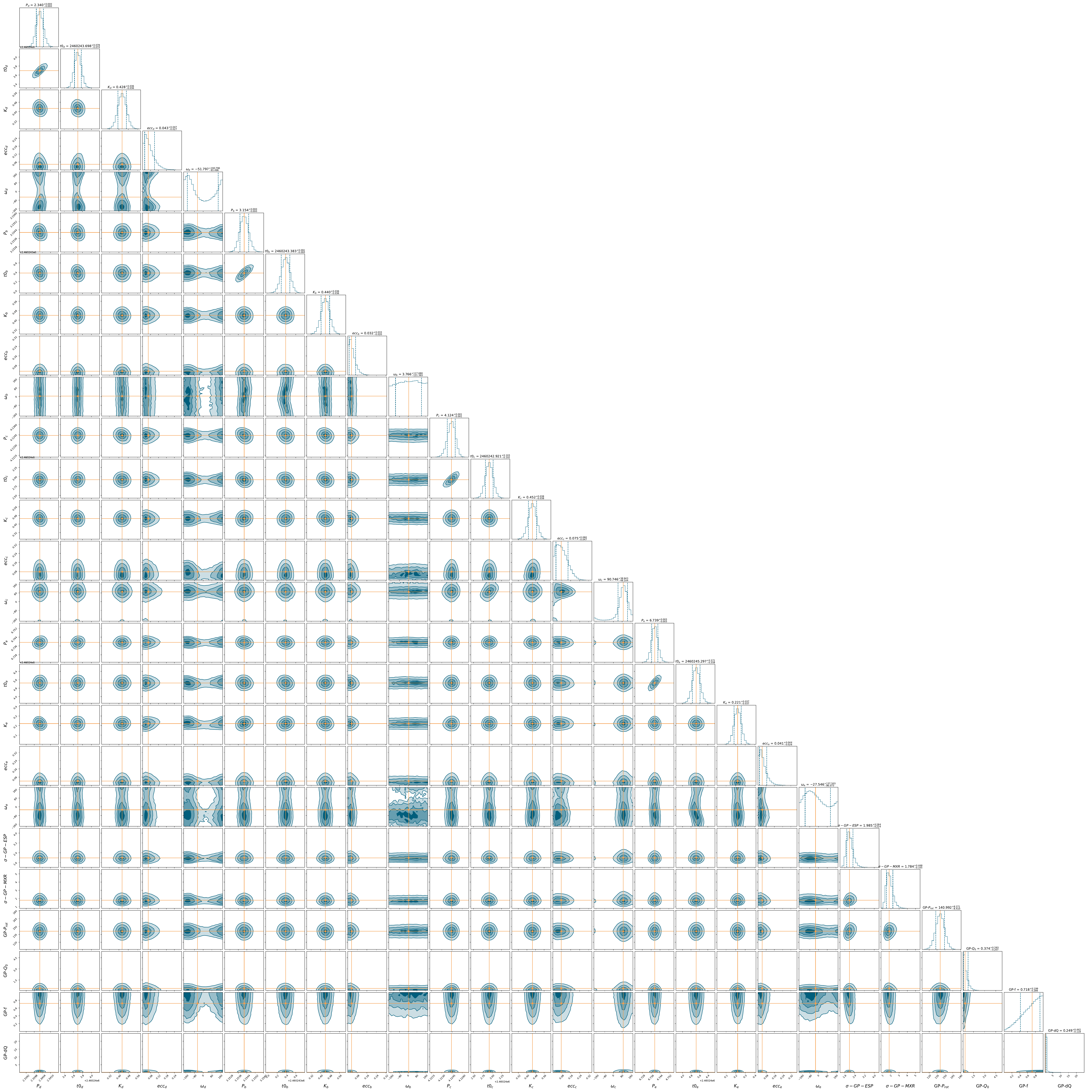} % Adjust the size to fit within the page
    \caption{Corner plot for the 4-Planet joint fit to MAROON-X Red channel and ESPRESSO datasets. We only plot this for the 20 planetary parameters and 6 GP hyperparameters.}
    \label{fig:corner}
\end{figure*}

%\begin{table}[]
%    \centering
%    \begin{tabular}{ccc}
%        \hline
%        \hline
%        Parameter & Posterior & Prior \\
%        \hline
%        
%        \hline
%    \end{tabular}
%    \caption{This table summarizes the posteriors and the priors for derived dataset parameters.}
%    \label{table:dataset}
%\end{table}


%% Include this line if you are using the \textbf, \replaced, \deleted

\end{document}